\documentclass[useAMS,usenatbib]{mn2e}
\usepackage{savesym}
\usepackage{txfonts}
\savesymbol{iint}
\savesymbol{iiint}
\savesymbol{iiiint}
\savesymbol{idotsint}
\usepackage{pdflscape}
\usepackage{graphicx}
\usepackage{longtable}
\usepackage{amssymb}
\usepackage[below]{placeins}
\usepackage{txfonts}
\usepackage{natbib}
\usepackage{amsmath}
\usepackage{float}
\restoresymbol{TXF}{iint}
\restoresymbol{TXF}{iiint}
\restoresymbol{TXF}{iiiint}
\restoresymbol{TXF}{idotsint}
\usepackage{multirow}
\usepackage{array}
\usepackage{xcolor}
\usepackage[percent]{overpic}
\bibpunct{(}{)}{;}{a}{}{,}

\def \rosat{{\emph{ROSAT}}}
\def \chandra{{\emph{Chandra}}}

\def \suzaku{{\emph{Suzaku}}}

\def\spose#1{\hbox to 0pt{#1\hss}}
\def\approxlt{\mathrel{\spose{\lower 3pt\hbox{$\sim$}}
        \raise 2.0pt\hbox{$<$}}}
\def\approxgt{\mathrel{\spose{\lower 3pt\hbox{$\sim$}}
        \raise 2.0pt\hbox{$>$}}}
\def\approxpropto{\mathrel{\spose{\lower 3pt\hbox{$\sim$}}
        \raise 2.0pt\hbox{$\propto$}}}
\mathchardef\twiddle="2218

\def\multleft#1{\hbox to size{\vbox {\halign {\lft{##}\cr #1}}\hfill}\par}
\def\multright#1{\hbox to size{\vbox {\halign {\rt{##}\cr #1}}\hfill}\par}

\def\today{\ifcase\month\or January\or February\or March\or April\or May\or
      June\or July\or August\or September\or October\or November\or December\fi
      \space\number\day, \number\year}
\def\<{\thinspace}

\def\ga{{\rm\thinspace gauss}}

\def\chandra{{\it Chandra}}

\def\rosat{{\it ROSAT}}

\makeatletter
\newcommand{\thickhline}{%
    \noalign {\ifnum 0=`}\fi \hrule height 1.2pt
    \futurelet \reserved@a \@xhline
}
\newcolumntype{"}{@{\hskip\tabcolsep\vrule width 1pt\hskip\tabcolsep}}
\makeatother

\bibpunct{(}{)}{;}{a}{}{,}

\title[Uniform metallicity in cluster outskirts]{A uniform metallicity in the outskirts of massive, nearby galaxy clusters}

\author[Urban et al.]{O. Urban$^{1,2}$, N. Werner$^{3,4,5}$\thanks{E-mail: wernernorbi@caesar.elte.hu}, S. W. Allen$^{1,2,6}$,  A. Simionescu$^{7}$, A. Mantz$^{1,2}$\\
$^1$Kavli Institute for Particle Astrophysics and Cosmology, Stanford University, 452 Lomita Mall, Stanford, CA 94305-4085, USA \\
$^2$Department of Physics, Stanford University, 382 Via Pueblo Mall, Stanford, CA 94305-4060, USA \\
$^3$MTA-E\"otv\"os University Lend\"ulet Hot Universe Research Group, P\'azm\'any P\'eter s\'et\'any 1/A, Budapest, 1117, Hungary \\
$^4$Department of Theoretical Physics and Astrophysics, Faculty of Science, Masaryk University, Kotl\'a\v{r}sk\'a 2, Brno, 611 37, Czech Republic \\
$^5$School of Science, Hiroshima University, 1-3-1 Kagamiyama, Higashi-Hiroshima 739-8526, Japan \\
$^6$SLAC National Accelerator Laboratory, 2575 Sand Hill Road, Menlo Park, CA 94025, USA\\
$^7$Institute of Space and Astronautical Science (ISAS), JAXA, 3-1-1 Yoshinodai, Chuo-ku, Sagamihara, Kanagawa, 252-5210, Japan
}

\begin{document}
\maketitle

\begin{abstract}
{\it Suzaku} measurements of a homogeneous metal distribution of $Z\sim0.3$ Solar in the outskirts of the nearby Perseus cluster suggest that chemical elements were deposited and mixed into the intergalactic medium before clusters formed, likely over 10 billion years ago. A key prediction of this early enrichment scenario is that the intracluster medium in all massive clusters should be uniformly enriched to a similar level. Here, we confirm this prediction by determining the iron abundances in the outskirts ($r>0.25r_{200}$) of a sample of ten other nearby galaxy clusters observed with {\it Suzaku} for which robust measurements based on the Fe-K lines can be made. Across our sample the iron abundances are consistent with a constant value, $Z_{\rm Fe}=0.316\pm0.012$~Solar ($\chi^2=28.85$ for 25 degrees of freedom). This is remarkably similar to the measurements for the Perseus cluster of $Z_{\rm Fe}=0.314\pm0.012$~Solar, using the Solar abundance scale of \citet{asplund2009}.

\end{abstract}


\section{Introduction}

Clusters of galaxies, the most massive objects in the Universe, are continuously growing, both by the steady accretion of matter from their surrounding environment, and by occasional mergers with smaller sub-clusters. The diffuse intergalactic gas accreted by clusters is rapidly shock heated, giving rise to the hot ($10^7$--$10^8$~K) X-ray emitting intra-cluster medium (ICM) that pervades clusters. The ICM is in approximate virial equilibrium, with the outer boundary of the virialized region - the virial radius - being approximately equal to $1.3r_{200}$, where within $r_{200}$ the mean enclosed mass density of the cluster is 200 times the critical density of the Universe at the cluster redshift \citep{lacey1993}. Galaxy clusters are also unique astrophysical laboratories that allow us to study nucleosynthesis and the chemical enrichment history of the Universe \citep[see][]{werner2008}. The deep gravitational potential wells of galaxy clusters hold all of the metals ever produced by stars in member galaxies, making them archaeological treasure troves to study the integrated history of star formation \citep{deplaa2007,mernier2016}. The dominant fraction of the metals in clusters currently resides within the hot ICM, which constitutes $\gtrsim 70$\% of the baryonic mass content for systems above $1.4\times10^{14}~M_{\odot}$ \citep{giodini2009}. However, when and how these metals were injected into the intergalactic medium is not well understood.

Most of the line emission from metals in the ICM arises from K- and L-shell transitions of highly ionized elements \citep[see][]{bohringer2010}. Because the ICM is in collisional ionization equilibrium and is optically thin, the equivalent widths of the emission lines can be converted directly into elemental abundances. The strongest line emission in the X-ray band is produced by the K-shell transitions of helium-like iron, making it an excellent tracer of chemical enrichment. 
	
It has been known for about 40 years that a significant portion of the hot plasma in the central regions of galaxy clusters (the inner $\sim0.3r_{200}$) has been enriched by iron produced in stars to about one-third to one-half of the Solar value \citep{mushotzky1978,mushotzky1981}. In the central regions of clusters with strongly peaked ICM density distributions the abundance of iron is also peaked \citep[e.g.][]{degrandi2004}, but decreases with radius to about one-third Solar \citep[assuming the Solar abundances of][]{asplund2009} beyond about $0.2r_{200}$ \citep{leccardi2008}. Due to the low X-ray surface brightness in the outskirts of clusters, metal abundance measurements beyond one-half of the virial radius of clusters remain sparse.

The best measurements of the Fe abundance distribution at large radii were performed using the \suzaku\ Key Project data (1 Ms observation along 8 azimuthal directions) of the Perseus cluster, which provided 78 data points outside of the cluster core ($r>0.25r_{200}$). These data revealed a remarkably uniform iron abundance, as a function of radius and azimuth, that is statistically consistent with a constant value of $Z_{\rm Fe} = 0.314 \pm 0.012$ Solar \citep[using the Solar abundance scale of ][]{asplund2009} out to  $r_{200}$ \citep{werner2013}. Subsequent \suzaku\ observations of the Virgo cluster extended these measurements to elements other than iron indicating an uniform chemical composition throughout the cluster volume \citep{simionescu2015}. The observed homogeneous distribution suggests that most of the metal enrichment of the ICM occurred before the cluster formed and its entropy distribution became stratified, preventing further efficient mixing. A key prediction of this early enrichment scenario is that the ICM in all massive clusters should be uniformly enriched to a similar level  \citep{werner2013,fabjan2010,biffi2017}. 

In order to test these predictions, we have analysed all archival observations of nearby galaxy clusters observed with {\it Suzaku} for which data extend to $r\sim r_{200}$ and robust measurements based on the Fe-K lines can be performed at $r > 0.25r_{200}$. (Because the Fe-K complex is by far the strongest line complex in the X-ray spectrum, the word metallicity will in this paper refer to and will be used interchangeably with the Fe abundance.) Our sample spans a redshift range $z=0.017$--$0.183$ and a temperature range of about $2.5$--$9$~keV \citep[the corresponding range of virial masses is about 1.4--$14\times10^{14}~M_\odot$;][]{arnaud2005}. The selected temperature range permits metallicity measurement using the Fe-K lines, allowing us to largely avoid multi-temperature biases arising from the measurements of the Fe-L complex \citep{buote2000}.

Sect.~\ref{sect:metObsanal} describes the data analysis, spectral modeling, and the treatment of the X-ray background.  In~Sect.~\ref{sect:metResults}, we present the results. Finally, in Sect.~\ref{discussion} and \ref{sect:metConclusions}, we briefly discuss the implications of these results and draw our conclusions.

\begin{table}
\scriptsize
\centering
\caption{Details of \emph{Suzaku} observations used in the analysis. The
columns show, respectively, the target name, the \suzaku\ observation ID, the date of the
observation and the clean exposure time.}
\label{tab:metData}
\begin{tabular}{lccc}
\hline\hline
\textbf{Name}&\textbf{Obs. ID}&\textbf{Obs. date}&\textbf{Exposure (ks)}\\
\hline
ABELL 262 CENTER    &802001010&2007-08-17& 10.9\\ 
ABELL 262 OFFSET 1$^\dagger$  &802079010&2007-08-06& 11.0\\
ABELL 262 OFFSET 2  &802080010&2007-08-08& 14.6\\
NCG 669$^\dagger$   &804049010&2009-07-05& 50.3\\
A262 NE1            &808108010&2014-02-14& 8.4 \\
A262 NE2$^\dagger$  &808109010&2014-02-15& 1.4 \\
A262 NE3$^\dagger$  &808110010&2014-02-15& 9.8 \\
A262 NE4            &808111010&2014-02-16& 31.0\\ 
A262 E1             &808112010&2014-02-06& 9.2\\ 
A262 E2$^\dagger$   &808113010&2014-02-07& 5.4\\
A262 E3             &808114010&2014-02-13& 10.9\\
A262 E4$^\dagger$   &808115010&2014-02-13& 9.8\\ 
\hline
ABELL 1795$^\dagger$  &800012010&2005-12-10&13.1\\
ABELL 1795 Near North$^\dagger$ &800012020&2005-12-10&24.4\\
ABELL 1795 Far North  &800012030&2005-12-11&5.9\\
ABELL 1795 Near South &800012040&2005-12-11&11.2\\
ABELL 1795 Far South  &800012050&2005-12-12&12.5\\
A1795\_FAR\_NORTHEAST$^\dagger$ &804082010&2009-06-28&4.7\\
A1795\_FAR\_SOUTHWEST &804083010&2009-06-29&8.3\\
A1795\_FAR\_WEST      &804084010&2009-06-26&8.5\\
A1795\_NEAR\_WEST     &804085010&2009-06-27&4.1\\
\hline
A1689-OFFSET1         &803024010&2008-07-23&9.1\\
A1689-OFFSET2         &803025010&2008-07-24&7.5 \\
A1689-OFFSET3         &803026010&2008-07-25&8.9\\
A1689-OFFSET4$^\dagger$      &803027010&2008-07-26&6.2 \\
ABELL 1689 (OFFSET)$^\dagger$   &808089010&2013-06-27&17.5\\
ABELL 1689 (OFFSET)   &808089020&2013-06-30&6.4\\
ABELL 1689 (OFFSET)   &808089030&2013-12-31&53.2\\
ABELL 1689 (OFFSET)   &808089040&2014-01-13&7.1\\
\hline
HYDRA A-1$^\dagger$   &805007010&2010-11-08& 6.9   \\
HYDRA A-2             &805008010&2010-11-09& 6.9   \\
HYDRA A SE            &807087010&2012-06-07& 5.5   \\
HYDRA A FAR SE        &807088010&2012-06-04& 5.6   \\
HYDRA A FAR N         &807089010&2012-06-05& 5.6   \\
HYDRA A SW$^\dagger$  &807090010&2012-11-10& 39.8  \\
HYDRA A OUT           &807091010&2012-06-07& 0.3   \\
\hline
ABELL 2029$^\dagger$  &802060010&2008-01-08&27.6\\
A2029\_1              &804024010&2010-01-28&3.5\\
A2029\_2              &804024020&2010-01-28&2.6\\
A2029\_3              &804024030&2010-01-28&6.3\\
A2029\_4              &804024040&2010-01-29&1.9\\
A2029\_5$^\dagger$    &804024050&2010-01-30&3.5\\
\hline
A2142                 &801055010&2007-01-04&12.0   \\
A2142 OFFSET 1        &802030010&2007-08-04&7.1    \\
A2142 OFFSET 2        &802031010&2007-09-15&57.7   \\
A2142 OFFSET 3        &802032010&2007-08-29&7.0    \\
FILAMENT OF GALAXIES  &805029010&2010-07-29&19.9   \\
\hline
ABELL 2204            &801091010&2006-09-17&14.3   \\
A2204\_FIELD\_1$\dagger$       &805056010&2010-09-01&5.9    \\ 
A2204\_FIELD\_2       &805057010&2010-08-27&6.9    \\
A2204\_FIELD\_3       &805058010&2010-08-28&6.8    \\
\hline
A133\_W               &805019010&2010-06-07&50.0 \\
A133\_N               &805020010&2010-06-05&50.2 \\
A133\_E$^\dagger$     &805021010&2010-06-09&51.6 \\
A133\_S$^\dagger$     &805022010&2010-06-08&51.1 \\
A133\_FIELD\_1        &808081010&2013-12-19&53.6 \\
A133\_FIELD\_2        &808082010&2013-12-20&50.6 \\
A133\_FIELD\_3        &808083010&2013-12-05&51.9 \\
A133\_FIELD\_4$^\dagger$        &808084010&2013-12-06&52.5 \\
\hline
SWIFT J0250.7+4142    &709006010&2014-08-03&82.2 \\
AWM7$^\dagger$        &801035010&2006-08-07&19.0 \\
AWM7 EAST OFFSET      &801036010&2006-08-05&38.5 \\
AWM7 WEST OFFSET$^\dagger$      &801037010&2006-08-06&39.8 \\
AWM7 EAST OFFSET      &802044010&2008-01-27&85.6 \\
AWM7 SOUTH OFFSET$^\dagger$     &802045010&2008-01-29&31.3 \\ 
AWM7 SOUTH OFFSET$^\dagger$     &802045020&2008-02-23&91.2 \\
AWM7 45' EAST         &806008010&2011-08-07&36.9 \\
AWM7 27' SOUTH$^\dagger$        &806009010&2012-02-18&35.0 \\
AWM7 45' SOUTH        &806010010&2012-02-17&34.4 \\
AWM7 NW1              &808023010&2014-02-17&14.9 \\
AWM7 NW2$^\dagger$              &808024010&2014-02-17&35.3 \\
AWM7 SE1              &808025010&2014-02-18&16.8 \\
AWM7 SE2$^\dagger$              &808026010&2014-02-19&35.3 \\
\hline\hline   
\multicolumn{4}{l}{$^\dagger$observations influenced by SWCX}
\end{tabular}
\end{table}

\begin{table*}
\centering   
\caption{Central coordinates, redshifts, mean temperatures, and values of $r_{200}$ for the clusters in our sample. The central coordinates are from the MCXC \citep{piffaretti2011}, except for A~1689, which are from the RASS-BSC \citep{voges1999}.}
\label{tab:metClusters}
\begin{tabular}{l|llccc}
\hline\hline
&RA (J2000)& dec (J2000)& $z$&$kT_{\rm ref}^{\dagger}$ (keV)&$r_{200, {\rm ref}}\,({\rm Mpc})$\\
\hline
\textbf{A~262}   &01h 52m 46.8s &+36$^{\circ}$ 09' 05''   &0.017&2.3& 1.52 \citep{neill2001}\\
\textbf{A~1795}  &13h 48m 53.0s &+26$^{\circ}$ 35' 44''   &0.063&6.2& 1.90 \citep{bautz2009}\\  
\textbf{A~1689}  &13h 11m 29.5s &$-$01$^{\circ}$ 20' 14'' &0.183&8.6& 2.50 \citep{umetsu2008}\\ 
\textbf{Hydra~A} &09h 18m 06.5s &$-$12$^{\circ}$ 05' 36'' &0.054&3.8& 1.48 \citep{sato2012}\\  
\textbf{A~2029}  &15h 10m 55.0s &+05$^{\circ}$ 43' 14''   &0.077&7.9& 1.92 \citep{walker2012b}\\
\textbf{A~2204}  &16h 32m 46.5s &+05$^{\circ}$ 34' 14''   &0.152&6.4& 1.84 \citep{reiprich2009}\\
\textbf{A~2142}  &15h 58m 20.6s &+27$^{\circ}$ 13' 37''   &0.091&8.5& 2.48 \citep{akamatsu2011}\\
\textbf{A~133}   &01h 02m 42.1s &$-$21$^{\circ}$ 52' 25'' &0.057&4.0& 1.60 \citep{morandi2014}\\  
\textbf{AWM~7}   &02h 54m 29.5s &+41$^{\circ}$ 34' 18''   &0.017&3.7& 1.47 \citep{walker2014}\\ 
\hline\hline
\multicolumn{4}{l}{$^\dagger$\citet{ikebe2002}}  
\end{tabular}
\end{table*}

\section{Observations and Data Analysis}
\label{sect:metObsanal}

The details of the \suzaku\ observations for each of the 9~clusters
analyzed in this study are shown in~Tab.~\ref{tab:metData}.  For each
cluster we analyzed the data from all available X-Ray Imaging Spectrometers
(XIS 0, 1, 2\footnote{XIS2 was lost to a likely micrometeoroid hit on 2006
November~9, and therefore its data are available only for the observations
from before this date.}, 3).

\subsection{Data Reduction}
\label{subs:metDataReduction}

We obtained the initial cleaned event lists using the standard criteria
provided by the XIS~team\footnote{Arida,~M., XIS Data Analysis,\\
http://heasarc.gsfc.nasa.gov/docs/suzaku/analysis/abc/node9.html}.  There is
a gradual increase in the number of flickering pixels in the XIS~detectors
with time, which may affect the measurements, if unaccounted for.  We used
maps provided by the XIS~team\footnote{The current maps, as well as the recipe for the removal
process, are available at
http://www.astro.isas.ac.jp/suzaku/analysis/xis/nxb\_new/} to remove the flickering pixels from the cluster
observations, as well as from the night Earth observations, which were later used
 to create the non-X-ray background (NXB) data products.

We checked for likely solar wind charge-exchange (SWCX) emission
contamination using the WIND Solar Wind Experiment data\footnote{The
data are available at: ftp://space.mit.edu/pub/plasma/wind/kp\_files/},
following the analysis of~\citet{fujimoto2007}.  In the case of affected
pointings, marked in~Tab.~\ref{tab:metData}, we only used data above
1.5~keV, where no SWCX emission lines are expected.

We filtered out times of low geomagnetic cut-off rigidity (COR$>$6~GV). 
Ray-tracing simulations of spatially uniform extended emission were used to
perform vignetting corrections \citep{ishisaki2007}.  For the XIS~1 data
obtained after the reported charge injection level increase on
2011~June~1st, we have excluded two adjacent rows on either side of the
charge-injected rows. (The standard is to exclude one row on either side.)

\subsection{Image analysis}
\label{subs:metImaging}

We extracted images from all XIS~detectors in the $0.7-7.0$~keV energy band, removing $\sim30$~arcsec regions around the edges.  We extracted
instrumental background images in the same energy band from
flickering-pixel-subtracted night Earth observations using the tool
\textsc{xisnxbgen}. We subtracted the background images from the cluster
images before applying the vignetting correction.  The resulting mosaics for
each cluster are shown
in~Figs.~\ref{fig:a262portrait}--\ref{fig:awm7portrait}.

\subsection{Point Source Detection}
\label{subs:metPsources}

The initial identification of the point sources was carried out using the
\textsc{ciao} tool \textsc{wavdetect}.  We used a single wavelet radius of
1~arcmin, which is approximately matched to the half-power radius of the X-ray telescopes on
\suzaku.  For each cluster we created a candidate set of point
sources assuming a source with a radius of 1~arcmin at each of the positions
identified by \textsc{wavdetect}. We then calculated X-ray
surface brightness profiles centered on the coordinates in
Tab.~\ref{tab:metClusters}, excluding the candidate set of point sources, 
and fitted an isotropic $\beta$-model to the surface
brightness profile of each cluster:

\begin{equation}
S_X=S_0\left[1+\left(\frac{r}{r_c}\right)^2\right]^{(-3\beta+0.5)}+S_{X, \rm
bkg}, 
\label{eqn:beta} 
\end{equation} 
where $r$ is the distance from the cluster center and the free parameters
are the normalization $S_0$, the core radius $r_c$ and $\beta$.  $S_{X, \rm bkg}$ is
the surface brightness of the X-ray background, which is assumed to be
constant across the whole area of the cluster.  The best-fit parameters for
the individual clusters are shown in~Tab.~\ref{tab:metSx}.

\begin{table}
\centering
\caption{$\beta$ model parameters for each cluster. See Eqn.~\ref{eqn:beta}
for the interpretation of the individual columns. Parameters $S_0$ and $S_{X, \rm bkg}$
are in the units of counts/s/pixel$^2$ and the core radius $r_c$ is in \suzaku\
pixels, where $1\,\text{arcmin}=7.2\,\text{pixels}$.}
\label{tab:metSx}
\begin{tabular}{l|cccc}
\hline\hline
&$S_0\times10^{5}$&$r_c$&$\beta$&$S_{X, \rm bkg}\times10^7$\\
\hline
\textbf{A~262}    &$1.45\pm0.05$&$18.3\pm0.7$&$0.51\pm0.01$&$0.98\pm0.035$\\
\textbf{A~1795}   &$7.78\pm0.26$&$14.7\pm0.4$&$0.70\pm0.01$&$0.90\pm0.04$\\
\textbf{A~1689}   &$4.40\pm0.43$&$10.6\pm0.8$&$0.78\pm0.02$&$1.18\pm0.03$\\
\textbf{Hydra~A}  &$3.88\pm0.23$&$14.1\pm0.8$&$0.67\pm0.02$&$1.03\pm0.06$\\
\textbf{A~2029}   &$9.52\pm0.37$&$14.1\pm0.5$&$0.69\pm0.01$&$1.37\pm0.04$\\
\textbf{A~2204}   &$4.44\pm0.17$&$16.5\pm0.7$&$0.93\pm0.03$&$2.89\pm0.08$\\
\textbf{A~2142}   &$4.58\pm0.11$&$20.3\pm0.6$&$0.70\pm0.01$&$8.15\pm0.05$\\
\textbf{A~133}    &$0.96\pm0.07$&$13.3\pm0.9$&$0.73\pm0.02$&$0.71\pm0.04$\\
\textbf{AWM~7}    &$2.37\pm0.05$&$29.4\pm0.8$&$0.55\pm0.01$&$0.75\pm0.04$\\
\hline\hline
\end{tabular}
\end{table}

We divided the mosaic images of the individual clusters by the best-fit
surface brightness models and used the resulting residual images to identify
by eye sources with radii larger than 1~arcmin.  In these cases, we manually
increased the sizes of the sources in question by the appropriate amount. 
The resulting updated sets of point sources (including substructures and artifacts that can appear at chip edges), which were excluded from the
subsequent spectral analysis, are shown with magenta circles in
Figs~\ref{fig:a262portrait} -- \ref{fig:awm7portrait}.

\subsection{Spectral Analysis}
\label{subs:metSpectral}

For each cluster, we extracted spectra from a series of concentric annular
regions centered on the respective cluster's centre (see
Tab.~\ref{tab:metClusters}). The width of the annuli was set to be at least
3~arcmin, with each annulus containing at least 3500~cluster counts, allowing us, in principle, to
measure the Fe~abundance with a relative uncertainty of at most $20\%$. 
The resulting annuli are shown in yellow
in~Figs.~\ref{fig:a262portrait}--\ref{fig:awm7portrait}.
Instrumental background spectra were created using Night Earth observations. 

We rebinned each spectrum to a minimum of one count per bin,
employing the extended C-statistic \citep{cash1979,arnaud1996} in the
fitting.  For each spectrum, we constructed an individual response matrix
with a resolution of 16~eV, spanning the $0.2-9.5$~keV energy band\footnote{These
rebinned matrices require $\sim25$~times less disk space compared to the default
choice of 2~eV resolution and the full $0.2-16.0$~keV \suzaku\ energy band,
allowing us to model all the spectra for a given cluster simultaneously and
speeding up the analysis, without compromising the accuracy of the results.}.
We used the task \textsc{xissimarfgen} (version 2010-11-05) to create
ancillary response files (assuming uniform emission from a circular region
with the radius of 20~arcmin).

We used \textsc{xspec} \citep[][version 12.9.0]{arnaud1996} to model the
spectra.  For each cluster we modeled all spectra simultaneously, using the
$0.7-7.0$~keV band for the front illuminated XIS~0, XIS~2 and
XIS~3 detectors, and the $0.6-7.0$~keV band for the back illuminated XIS~1, except for observations with possible SWCX contamination (see
Tab.~\ref{tab:metData}), where we used the $1.5-7.0$~keV energy band.  We
modeled the ICM emission in each annulus as a single temperature plasma in collisional
ionisation equilibrium using the absorbed \textsc{apec (ATOMDB 3.0.3)} model
\citep{smith2001}. 
													
For a given annulus, we used a single temperature and metallicity.  Normalizations were allowed to vary among individual observations,
but were tied among the detectors in a single observation; in other
words, all spectra from a given observation were members of a single fitting
group in \textsc{xspec}.  We used the abundance table of~\citet{asplund2009}
in the analysis. Galactic absorption was set to the average column
along the line of sight inferred from the Leiden/Argentine/Bonn Survey
\citep{kalberla2005}. The uncertainties in
all derived parameters were determined using Markov Chain Monte Carlo (MCMC)
simulations. 
After removing the burn in period and thinning each chain, we used the mean and the standard deviation as the value and the uncertainty of each derived parameter, respectively.

\begin{table*}
\setlength{\extrarowheight}{4pt}
\centering
\caption{The CXFB model parameters for the individual clusters. The four
CXFB model components we used are the power-law component (PL), the Galactic
halo (GH), the hot foreground component (HF) and the Local Hot Bubble (LHB).
Subscript n stands for normalization, kT for temperature and ind for index.
Normalizations are in units of $\int n_en_H\,dV\times\frac{10^{-14}}{4\pi\left[D_A(1+z)\right]^2}\frac1{20^2\pi}\,\text{cm}^{-5}\,\text{arcmin}^{-2}$.}
\label{tab:metCxfb}
\begin{tabular}{l|cccccccc}
\hline\hline
&PL\textsubscript{ind}&PL\textsubscript{n}$\times10^{4}$&GH\textsubscript{kT}&GH\textsubscript{n}$\times10^3$&HF\textsubscript{kT}&HF\textsubscript{n}$\times10^4$&LHB\textsubscript{kT}&LHB\textsubscript{n}$\times10^4$\\
\hline
\textbf{A~262}    &$1.43_{-0.05}^{+0.05}$&$9.38_{-0.64}^{+0.67}$&$0.18_{-0.01}^{+0.01}$&$2.82_{-0.19}^{+0.13}$&$0.94_{-0.05}^{+0.04}$&$2.59_{-0.22}^{+0.12}$&$0.103_{-0.002}^{+0.002}$&$9.01_{-0.30}^{+0.18}$\\
\textbf{A~1795}   &$1.38_{-0.08}^{+0.05}$&$10.07_{-0.79}^{+0.84}$&$0.22_{-0.01}^{+0.01}$&$0.91_{-0.08}^{+0.18}$&N/A&N/A&$0.10_{-0.01}^{+0.01}$&$49.10_{-0.36}^{+0.26}$\\
\textbf{A~1689}   &$1.34_{-0.03}^{+0.04}$&$9.16_{-0.41}^{+0.47}$&$0.18_{-0.02}^{+0.01}$&$2.43_{-0.40}^{+0.50}$&$0.59_{-0.03}^{+0.05}$&$2.11_{-0.32}^{+0.64}$&$0.10_{-0.01}^{+0.01}$&$13.65_{-0.20}^{+0.23}$\\
\textbf{Hydra~A}  &$1.39_{-0.09}^{+0.07}$&$8.61_{-0.90}^{+1.15}$&$0.12_{-0.01}^{+0.01}$&$4.45_{-1.30}^{+1.40}$&$0.79_{-0.16}^{+0.11}$&$1.91_{-0.16}^{+0.36}$&$0.10_{-0.01}^{+0.01}$&$9.53_{-1.20}^{+1.20}$\\
\textbf{A~2029}   &$1.25_{-0.08}^{+0.07}$&$7.01_{-0.63}^{+0.69}$&$0.18_{-0.02}^{+0.01}$&$6.18_{-1.05}^{+1.15}$&$0.58_{-0.01}^{+0.02}$&$13.95_{-1.30}^{+0.90}$&$0.10_{-0.01}^{+0.01}$&$13.4_{-0.37}^{+0.19}$\\
\textbf{A~2204}   &$1.49_{-0.10}^{+0.10}$&$12.52_{-1.47}^{+1.62}$&$0.227_{-0.005}^{+0.008}$&$10.63_{-2.25}^{+0.45}$&$0.61_{-0.02}^{+0.02}$&$10.25_{-0.12}^{+0.22}$&$0.13_{-0.01}^{+0.06}$&$17.50_{-2.00}^{+1.08}$\\
\textbf{A~2142}   &$1.37_{-0.05}^{+0.06}$&$9.19_{-0.61}^{+0.64}$&$0.146_{-0.004}^{+0.004}$&$4.88_{-0.55}^{+55}$&$0.64_{0.02}^{+0.06}$&$2.51_{-0.22}^{+0.14}$&$0.09_{-0.01}^{+0.01}$&$7.33_{0.30}^{+0.60}$\\
\textbf{A~133}    &$1.54_{-0.07}^{+0.10}$&$10.37_{-0.95}^{+1.27}$&$0.147_{-0.006}^{+0.012}$&$1.93_{-0.20}^{+0.58}$&$0.61_{-0.16}^{+0.06}$&$1.18_{-0.20}^{+0.35}$&$0.08_{-0.01}^{+0.01}$&$7.33_{-0.90}^{+0.65}$\\
\textbf{AWM~7}    &$1.52_{-0.05}^{+0.02}$&$11.5_{-0.72}^{+0.30}$&$0.150_{-0.002}^{+0.004}$&$3.60_{-0.23}^{+0.13}$&$0.78_{-0.02}^{+0.02}$&$4.16_{-0.12}^{+0.14}$&$0.09_{-0.01}^{+0.01}$&$9.65_{-0.07}^{+0.10}$\\
\hline\hline
\end{tabular}
\end{table*}

\subsection{Modeling the X-ray Foreground and Background}
\label{subs:metCxfb}

At large clustercentric radii, the cosmic X-ray foreground/background (CXFB)
makes up a dominant fraction of the total X-ray emission, requiring careful
modeling. 
Our spectral model for the CXFB included four components -- an
absorbed power law (PL) due to the unresolved point sources
\citep{deluca2004}, an absorbed thermal component modeling the Galactic
halo emission \citep[GH,][]{kuntz2000}, a potential 0.6~keV foreground component
that we will from now on refer to as the hot foreground
\citep[HF,][]{masui2009,yoshino2009}, and an unabsorbed thermal component
modeling the emission from the local hot bubble \citep[LHB,][]{sidher1996}.

In order to better constrain the low-temperature CXFB components, we used
the X-Ray Background
Tool\footnote{http://heasarc.gsfc.nasa.gov/cgi-bin/Tools/xraybg/xraybg.pl},
which calculates the average X-ray background spectra from the ROSAT All-Sky
Survey diffuse background maps.  For each cluster we obtained spectra from
six independent circular regions with radii $R=1.3r_{200}$ evenly
surrounding the cluster so that each touches the outer edge of a circle with radius $r=1.3r_{200}$ centered on the cluster core, as well as two neighbouring background regions. The distance from the cluster core ensures that the
spectra are not significantly contaminated by emission from the ICM.  This
setup, as opposed to using a single annular region, allowed us to
assign separate absorptions to each of the six regions, which can
potentially significantly influence the modeling at \rosat\ energies,
$0.7-2.0$~keV.

\begin{figure*}
\centering
\includegraphics[width=.6\textwidth]{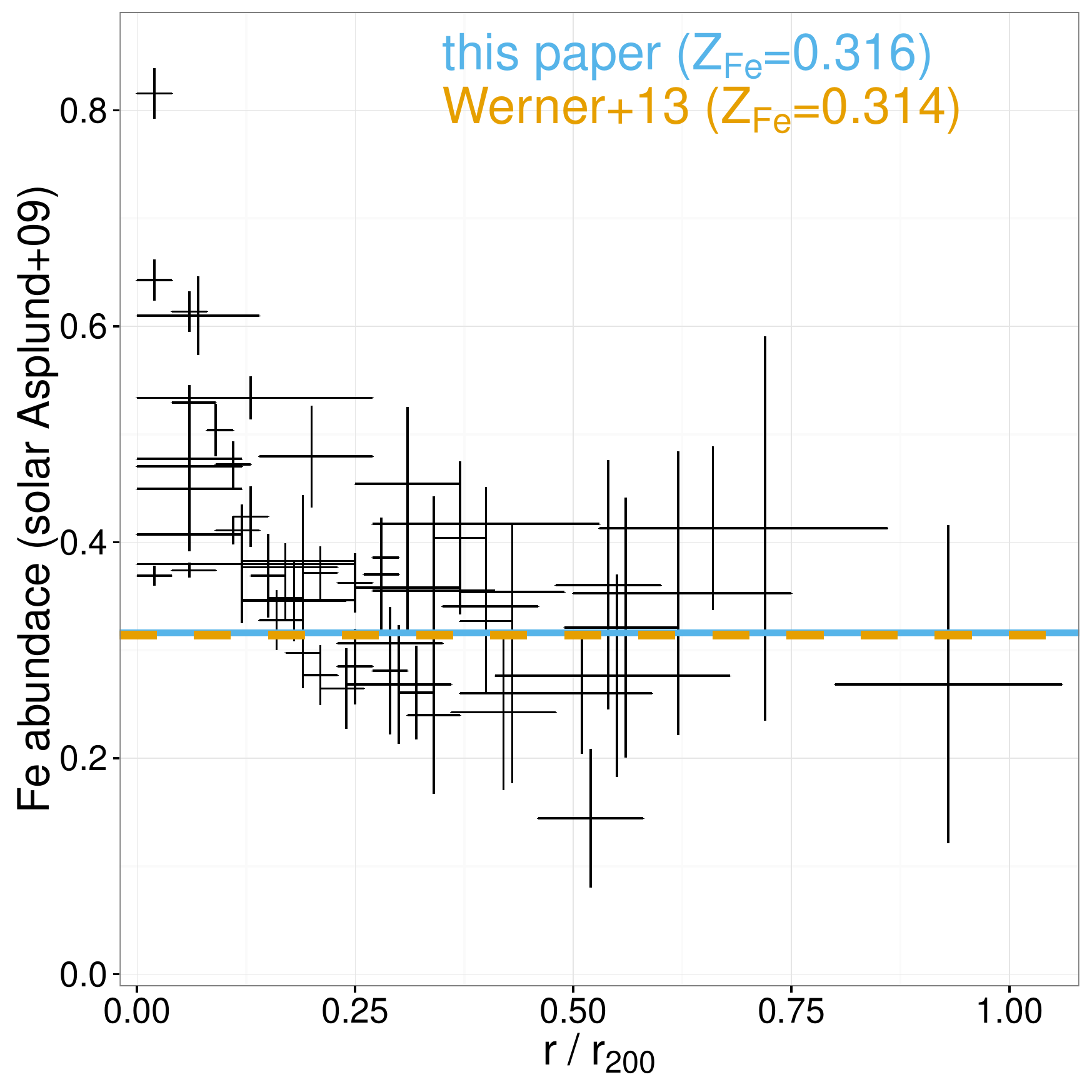}
\caption{Iron abundance measurements in our cluster sample plotted as a function of radius scaled to $r_{200}$. On average, the iron abundances peak in the cores of the clusters and decrease as a function of radius, flattening at radii $r>0.25r_{200}$. The average metallicity is shown as blue solid line. The dashed line shows the best fit metallicity reported by \citet{werner2013} for the Perseus cluster. }
\label{fig:radial}
\end{figure*}

As stated before, for each cluster we modeled all spectra simultaneously, including the CXFB model.  To limit the systematic effects that might
potentially influence the fit, we first separately determined the
PL~parameters, that were kept fixed during the subsequent modeling.  To do
this, we used the spectra from the outermost annulus in a given cluster in
the high energy band, $2.0-7.0$~keV, where the PL~component is dominant at
large clustercentric radii.  For six of the clusters in our sample,
A~133, A~1689, A~1795, A~2029, A~2142, and A~2204, the outermost
annulus covers only regions outside $r_{200}$. In these cases, we assumed
no significant cluster emission to contribute to the spectrum and therefore
the high energy fit included only the PL~model. For the remaining clusters,
A~262, AWM~7 and Hydra~A, we accounted for potential cluster emission by
including an \textsc{apec} component in the high energy fit, fixing its
temperature to $kT=2$~keV and keeping the
normalizations in the individual observations free. The best-fit
PL~parameters are shown in the first two columns of Tab.~\ref{tab:metCxfb}.

In the subsequent spectral modeling, the remaining CXFB parameters were
kept free and tied among all spectra from a given cluster, with the
exception of the metallicity and the redshift of the three thermal
components, which were fixed at unity and zero, respectively.

For each cluster, we tied
together the ICM~temperatures and Fe~abundances in the neighboring annuli
where required in order to obtain a statistically significant constraint. The final CXFB model parameters for the individual clusters
are listed in Tab.~\ref{tab:metCxfb}.

\begin{figure*}
\centering
\includegraphics[width=.6\textwidth]{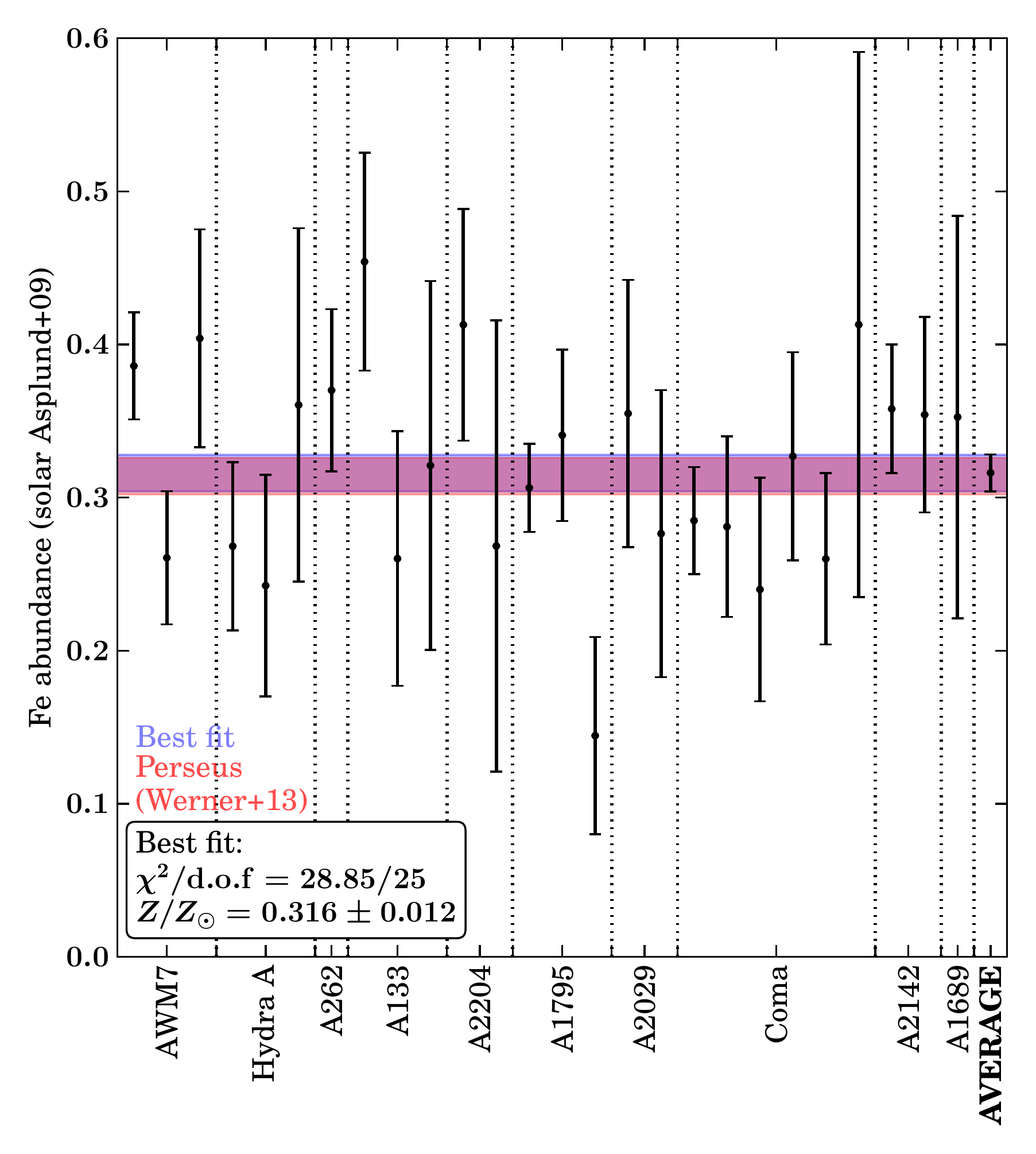}
\caption{Robust measurements (see text for details) of the iron abundances at $r>0.25r_{200}$ in
the individual clusters. The clusters have been ordered by mass from the
least to the most massive. The blue stripe marks the 68\% confidence interval
around the constant fit to these data, $Z_{\rm Fe}=0.316\pm0.012$~Solar. 
The red stripe shows the confidence interval around the best fit iron abundance 
reported by \citet{werner2013} for the Perseus cluster, $Z_{\rm Fe}=0.314\pm0.012$~Solar.}
\label{fig:metConservative}
\end{figure*}

\subsection{Criteria For Robust Metallicity Measurements}
\label{subs:metTrimming}

In each cluster we formally obtained profiles of temperature and metallicity
out to the outermost radii reached by the observations. However, due to the low
surface brightness of the ICM, the measurements at large radii may be
significantly influenced by systematic uncertainties, such as potential
variations of the CXFB model throughout the cluster. To address our main
scientific questions, it is therefore crucial to only use the metallicity
measurements which we are confident about. We used the following criteria to
identify these measurements:

\begin{itemize}
\item We only used the measurements at radii $r>0.25r_{200}$ to avoid the
central metallicity peak observed in most cool-core galaxy clusters. We used a
similar radial range in the Perseus cluster
\citep{werner2013}, where metallicity measurements inside $20'$
($r_{200}=82'$) were discarded. 
\item We only used annuli with ICM-signal-to-background (ISB) ratios
around the Fe-K lines higher than 10\%. This is defined as the ratio
of the total number of modeled counts received from the ICM to the sum
of modeled counts from the CXFB and the instrumental background, in a 1~keV
wide energy band centered on the appropriately redshifted Fe-K line (rest
energy $E=6.7\,\text{keV}$). The top right panels in
Figs.~\ref{fig:a262portrait}--\ref{fig:awm7portrait} show the profiles of
the ISB~ratios for the individual annuli (in blue), as well as the ratios
for the individual observations (in red).  The 10\% threshold is broadly
consistent with the measurements in the outermost regions of the Perseus
cluster \citep{werner2013,urban2014}.

\item Finally, we only used the annuli where the contamination from the neighbouring regions due to the wings of the broad point spread function (PSF) of the telescopes is small.   The half-power diameter (HPD) of the X-ray Telescopes (XRT) on board of \suzaku\ is
$\sim2'$, which causes a fraction of the emission from an object to be
registered elsewhere on the detector.  Addressing this issue is especially
important in the cool core clusters at relatively large distances
($z\gtrsim0.1$), since the emission from the metal-rich X-ray surface
brightness peak may bias spectral measurements out to larger radii. 
To test for this, we used \chandra\ surface brightness profiles with high
spatial resolution relative to \suzaku\ (binned to $\sim4''$), which we
convolved with a simple Gaussian model for the \suzaku\ point-spread
function with a HPD of 2~arcmin. Using this model, for each of our annuli
we calculated the fraction of emission that we expected to come from the
other annuli, and removed those where it exceeded 10\%.  Only the two most distant clusters in our sample, A~2204 and A~1689, were affected. For both systems, we removed the measurements
immediately outside $0.25r_{200}$ from the subsequent analysis.
\end{itemize}

\begin{table}
\begin{center}
\caption{Measurement radii and the best fit metallicities measured in our cluster sample. }
\begin{tabular}{lc|c}
\hline\hline
\textbf{cluster}&\textbf{r/r\textsubscript{200}}&\textbf{Z/Z}$_{\odot}$\\
\hline
\multirow{3}{*}{\textbf{AWM7}}&$0.285\pm0.015$&$0.386\pm0.035$\\
                              &$0.32\pm0.02$&$0.261\pm0.043$\\
                              &$0.37\pm0.03$&$0.404\pm0.071$\\
\hline
\multirow{3}{*}{\textbf{Hydra A}}&$0.30\pm0.06$&$0.268\pm0.055$\\
                                 &$0.42\pm0.06$&$0.243\pm0.072$\\
                                 &$0.54\pm0.06$&$0.361\pm0.115$\\
\hline
\textbf{A262}&$0.28\pm0.02$&$0.370\pm0.053$\\
\hline
\multirow{3}{*}{\textbf{A133}}&$0.31\pm0.06$&$0.454\pm0.071$\\
                              &$0.43\pm0.06$&$0.260\pm0.083$\\
                              &$0.56\pm0.07$&$0.321\pm0.120$\\
\hline
\multirow{2}{*}{\textbf{A2204}}&$0.66\pm0.13$&$0.413\pm0.076$\\
                               &$0.93\pm0.13$&$0.268\pm0.147$\\
\hline
\multirow{3}{*}{\textbf{A1795}}&$0.29\pm0.06$&$0.306\pm0.029$\\
                               &$0.40\pm0.05$&$0.341\pm0.056$\\
                               &$0.52\pm0.06$&$0.145\pm0.064$\\
\hline
\multirow{2}{*}{\textbf{A2029}}&$0.34\pm0.07$&$0.355\pm0.087$\\
                               &$0.55\pm0.14$&$0.276\pm0.094$\\
\hline
\multirow{2}{*}{\textbf{A2142}}&$0.31\pm0.06$&$0.358\pm0.042$\\
                               &$0.43\pm0.06$&$0.354\pm0.064$\\
\hline
\textbf{A1689}&$0.62\pm0.12$&$0.353\pm0.131$\\
\hline
\multirow{3}{*}{\textbf{Coma$^{\dagger}$}}
			     &$0.25\pm0.02$&$0.285\pm0.035$\\
                              &$0.29\pm0.02$&$0.281\pm0.059$\\
                              &$0.34\pm0.03$&$0.240\pm0.072$\\
			     &$0.40\pm0.03$&$0.327\pm0.068$\\
                              &$0.50\pm0.08$&$0.260\pm0.056$\\
                              &$0.71\pm0.13$&$0.41\pm0.18$\\
\hline\hline
\textbf{Average metallicity}&&$0.316\pm0.012$\\
\hline\hline
\end{tabular}\\
\end{center}
$^\dagger$\citet{simionescu2013}
\label{metaltable}
\end{table}

\section{Results}
\label{sect:metResults}

The best fit normalizations, temperatures and metallicities for the individual clusters are shown in the bottom panels of Figures \ref{fig:a262portrait}--\ref{fig:awm7portrait}. Most of the systems in our sample are so-called cooling core clusters with bright, relatively cool, metal-rich cores. To the iron abundance measurements in this work, we also added the iron abundances measured for the non-cool core Coma cluster by \citet{simionescu2013}.

Fig. \ref{fig:radial} shows all metallicity measurements in our cluster sample plotted as a function of radius scaled to $r_{200}$. The average metallicity (shown with the solid blue line) peaks in the central region and decreases as a function of radius, flattening at radii $r>0.25r_{200}$. 
We tested our results for biases associated with possible multi-temperature structure by fitting the data both in the full spectral band and above 2 keV. At radii $r>0.25r_{200}$ the two fits provided consistent results, indicating that cooler temperature components do not contribute significantly to the observed emission in the cluster outskirts.

After excluding the measurements at $r<0.25r_{200}$ and using all criteria outlined in the previous section, we are left with
26~individual metallicity measurements from 10~different clusters.  These measurements are shown in Table~\ref{metaltable} and Fig. \ref{fig:metConservative} where the clusters have been ordered by mass from the least to the most massive. There is no evidence for any trend in metallicity as a function of cluster mass. 
The measurements are consistent with being constant at $Z_{\rm Fe}=0.316\pm0.012$~Solar,
with $\chi^2=28.85$ for 25 degrees of freedom. This best fit value is statistically consistent with $Z_{\rm Fe}=0.314\pm0.012$ Solar reported for the Perseus cluster \citep{werner2013}, shown as a dashed line in Fig. \ref{fig:radial}.

\section{Discussion}
\label{discussion}

We find that across our sample of 10 clusters of galaxies the Fe abundances measured outside the central regions ($r>0.25r_{200}$) are consistent with a constant value, $Z=0.316\pm0.012$ Solar (Fig. \ref{fig:radial}). The metallicity measurements also show no significant trend with temperature (Fig. \ref{fig:metConservative}). 

Based on the uniform iron abundance distribution in the Perseus cluster, both as a function of radius and azimuth, statistically consistent with a constant value of $Z_{\rm Fe} = 0.314\pm0.012$ Solar out to $r_{200}$, \citet{werner2013} proposed that most of the metal enrichment of the intergalactic medium occurred before clusters formed, probably more than ten billion years ago ($z>2$), during the period of maximal star formation and black hole activity. A key prediction of the early enrichment scenario is that the ICM in all massive clusters should be uniformly enriched to a similar level. Previous indications for a uniform ICM enrichment include the small cluster to cluster scatter in the Fe abundance observed within $r_{500}$ \citep{matsushita2011,leccardi2008} and the observed pre-enrichment of the ICM between the clusters Abell 399/401 \citep{fujita2007}. Our observation of a constant iron abundance at large radii across a sample of 26 independent measurements for ten massive clusters  further confirms this early enrichment scenario. This early enrichment could have been driven by galactic winds \citep{deyoung1978} which would be strongest around the peak of star formation and AGN activity \citep[redshifts $z\sim2-3$][]{madau1996,brandt2005}.

Recent numerical simulations by \citet{fabjan2010} and \citet{biffi2017} indicate that while star-formation and supernova feedback are unable to enrich the intergalactic medium uniformly, simulations which also include feedback from AGN produce remarkably  homogeneous metallicity distribution in the ICM out to large radii. They show that the uniform metallicity is the result of a widespread
displacement of metal-rich gas by powerful AGN outbursts that occure {\it before} the epoch of maximal star-formation and AGN activity. \citet{biffi2017} conclude that  early AGN feedback acting on high-redshift ($z > 2$) small haloes, with shallow gravitational potential wells, was particularly efficient in spreading and mixing the metals. Given the complexity of the physics of the chemical enrichment processes, these simulation results should probably be considered tentative. However, our measurements provide an important anchor with which the results of these and future simulations can be compared, bringing more understanding into the process of chemical enrichment.  

The constant ratios of abundances of several elements observed throughout the Virgo cluster \citep{simionescu2015} as well as in the radial profiles of 44 clusters observed out to intermediate radii with {\it XMM-Newton} \citep{mernier2017} reveal that, during the early period of metal enrichment, the products of core-collapse and type Ia supernovae were well mixed. The estimated ratio between the number of SN Ia and the total number of supernovae enriching the ICM is about 15--20\%, generally consistent with the metal abundance patterns in our own Galaxy and only marginally lower than the SN Ia contribution estimated for the cluster cores \citep{simionescu2015}.

The most direct way to confirm the early enrichment scenario is to measure the core-excluded metallicity of clusters as a function of redshift. Contrary to initial findings \citep{Balestra2007,maughan2008,anderson2009,baldi2012}, under the early enrichment scenario, there should be no substantial redshift evolution in the ICM metallicity outside the central regions of clusters, out to $z \sim 2$. Recent results \citep{andreon2012,ettori2015,mcdonald2016, mantz2017} indicate that most metals in the ICM were already in place at $z = 1$, consistent with the picture of an early enrichment. 

At various overdensities, the chemical enrichment might proceed on different time scales or with different initial mass functions, resulting in a trend with cluster mass. Within the mass range probed by our sample (factor of $\sim10$), there is no evidence for dependence of ICM metallicity on total cluster mass. A more thorough analysis of the mass dependence will require reliable measurements of absolute abundances in low mass clusters and groups of galaxies, which are often made difficult for current CCD instruments by multi-temperature structure in the ICM \citep{simionescu2015,simionescu2017}. A lack of trend with cluster mass would either indicate a rate of metal enrichment in the early Universe that is independent of the density
contrast between different regions, or a very high efficiency of mixing on large scales.

If the ICM at large radii is clumpy and multiphase \citep{simionescu2011,urban2014,simionescu2017} then its best fit metallicity, derived using a single temperature model, might be biased \citep{avestruz2014}. The best fit Fe abundance is the most significantly biased at temperatures around 1~keV, where its value is determined based on the Fe-L lines, which are very sensitive to the underlying temperature structure \citep[see][]{buote2000}. The metallicities of the clusters in our sample are determined using the Fe-K lines and depending on the temperature structure could be biased by at most 30 per cent \citep[both toward higher and lower values;][]{rasia2008,simionescu2009b,gastaldello2010}. The fact that the spectral fits in the full band and above 2~keV give consistent results (see Section~\ref{sect:metResults}) indicates that if substantially cooler, denser clumps are present in the ICM, they do not contribute significantly to the observed emission measure and the metal budget. 

In the near future, the metallicities of groups and cooler clusters could be further studied  with the {\it Astrosat}  satellite \citep{singh2014}. The low earth orbit and the small inclination of the orbital plane of  {\it Astrosat}  provide a low and stable background environment that is required for cluster outskirts studies. The large field of view provides a sufficient grasp, enabling mapping the faint X-ray emission in the outskirts of nearby clusters that span large angular scales in the sky. Such observations will further test the possible mass-dependence of metallicity. Deep observations with {\it XMM-Newton} and {\it Chandra} will allow us to precisely determine the metallicity outside the cores of high redshift clusters, providing further constraints on the redshift evolution of metallicity. Observations with high spectral resolution obtained with the {\it X-ray Astronomy Recovery Mission (XARM)} will allow more accurately measured relative abundance ratios for clusters at low redshifts, testing our models of nucleosynthesis. In the further future, missions like {\it Athena} \citep{nandra2013} or {\it Lynx}\footnote{https://wwwastro.msfc.nasa.gov/lynx/} will allow detailed studies of metal abundances in high redshift clusters, providing comprehensive understanding of the metal cycle in the Universe.

\section{Conclusions}
\label{sect:metConclusions}

Here, we report 26~independent
metallicity measurements in the outskirts ($r>0.25r_{200}$) of ten nearby galaxy clusters. 
These measurements are consistent with a constant
value $Z_{\rm Fe}=0.316\pm0.012$~Solar. No significant trend of metallicity versus temperature or mass is observed. 

Our results corroborate the conclusions drawn from previous metallicity measurements at large radii in the Perseus cluster
\citep[][]{werner2013}. In particular, they confirm the predictions of an early
enrichment scenario, where the majority of metal enrichment occurs before
the cluster formation, at $z>2$.

\section*{Acknowledgments}
This work was supported in part by NASA grants NNX12AE05G and NNX13AI49G, and by the US Department of Energy under contract number DE-AC02-76SF00515, as well as by the Lend\"ulet LP2016-11 grant awarded by the Hungarian Academy of Sciences. The authors thank the {\it Suzaku} operation team and Guest Observer Facility, supported by JAXA and NASA.

\bibliographystyle{mnras}
\bibliography{clusters}

\appendix

\section{Individual clusters}

\begin{figure*} 
\begin{minipage}{.49\textwidth}
\includegraphics[width=\textwidth]{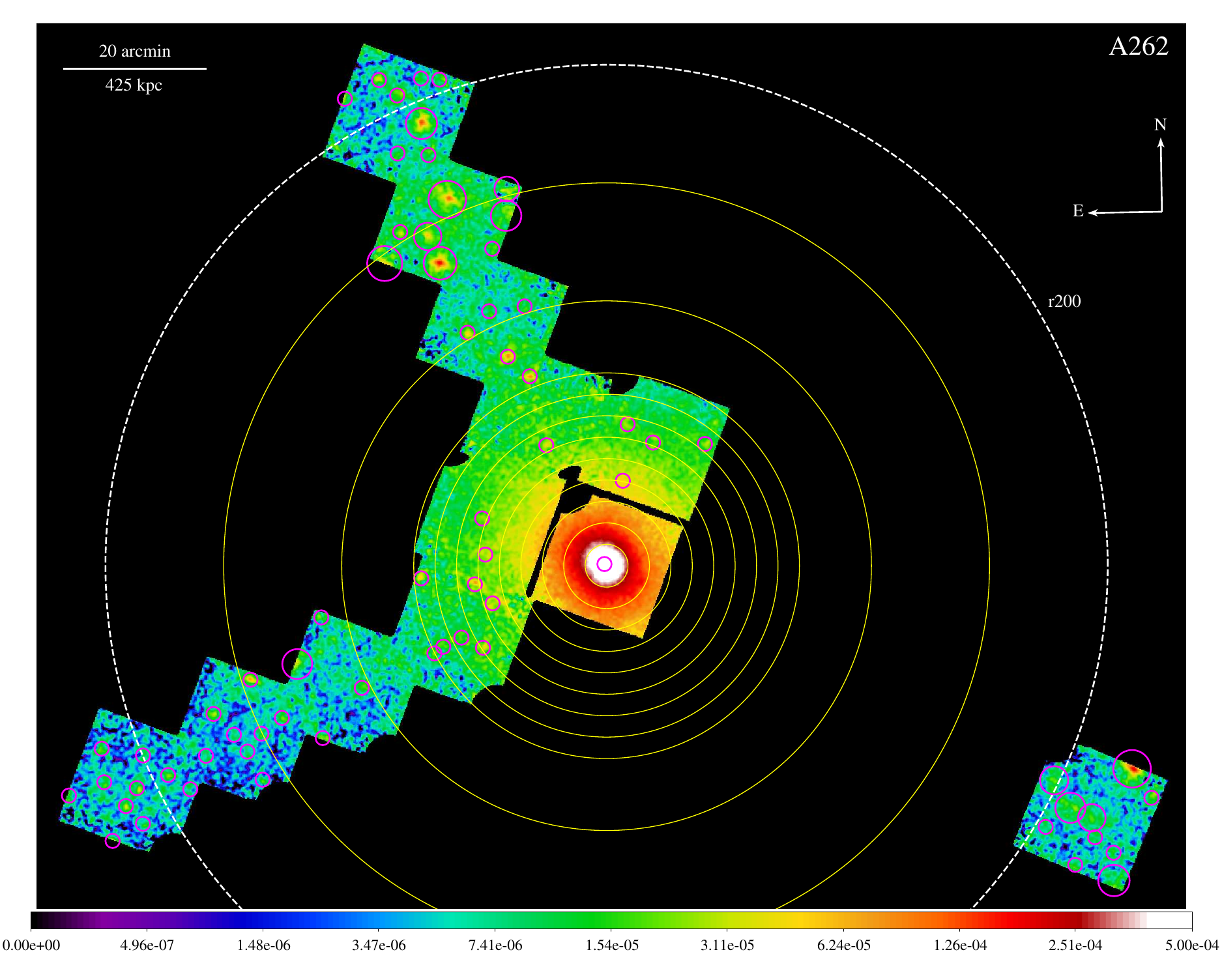} 
\end{minipage}
\begin{minipage}{.49\textwidth}
\includegraphics[width=\textwidth]{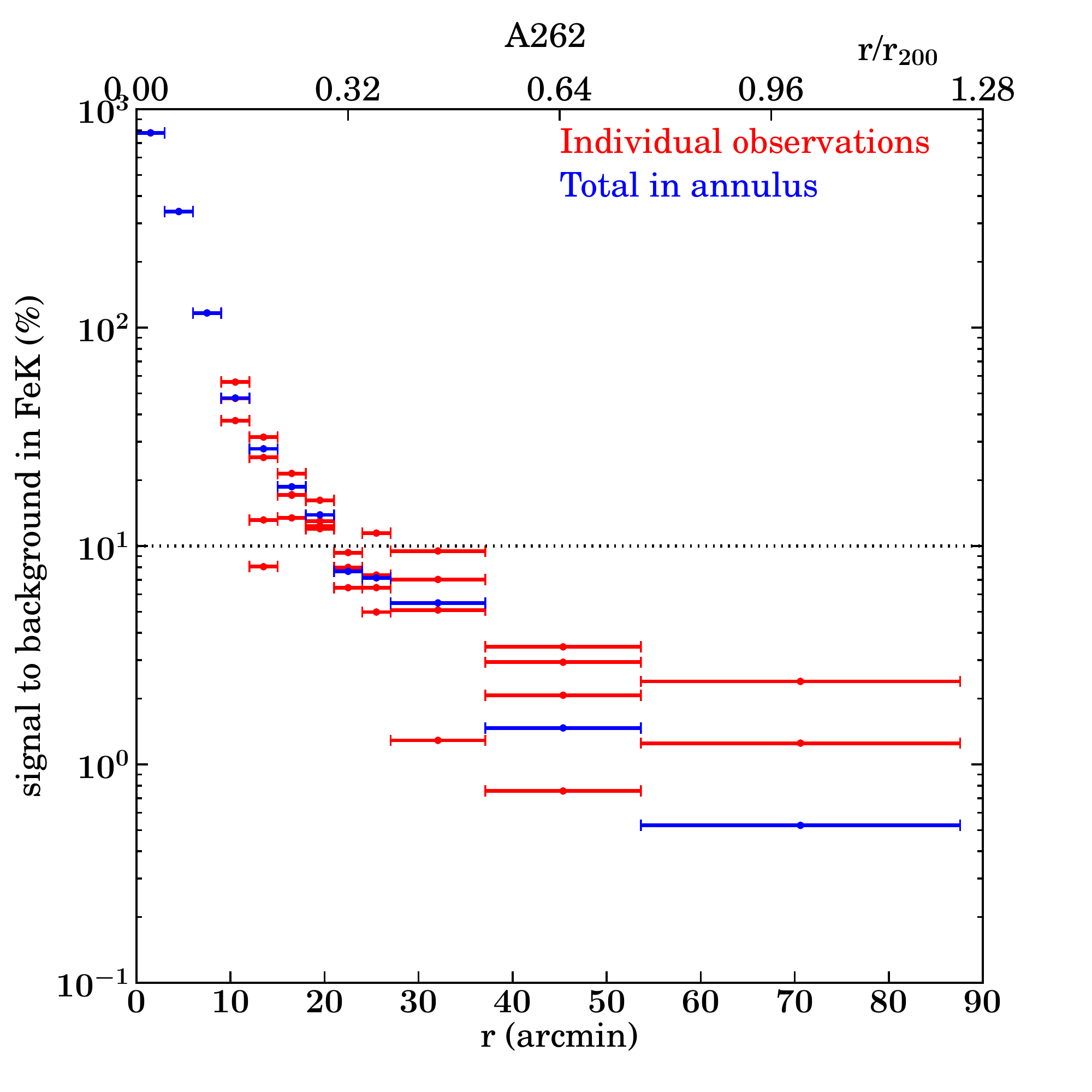}
\end{minipage} 
\begin{minipage}{.46\textwidth}
\vspace{-0.5cm}
\includegraphics[width=\textwidth]{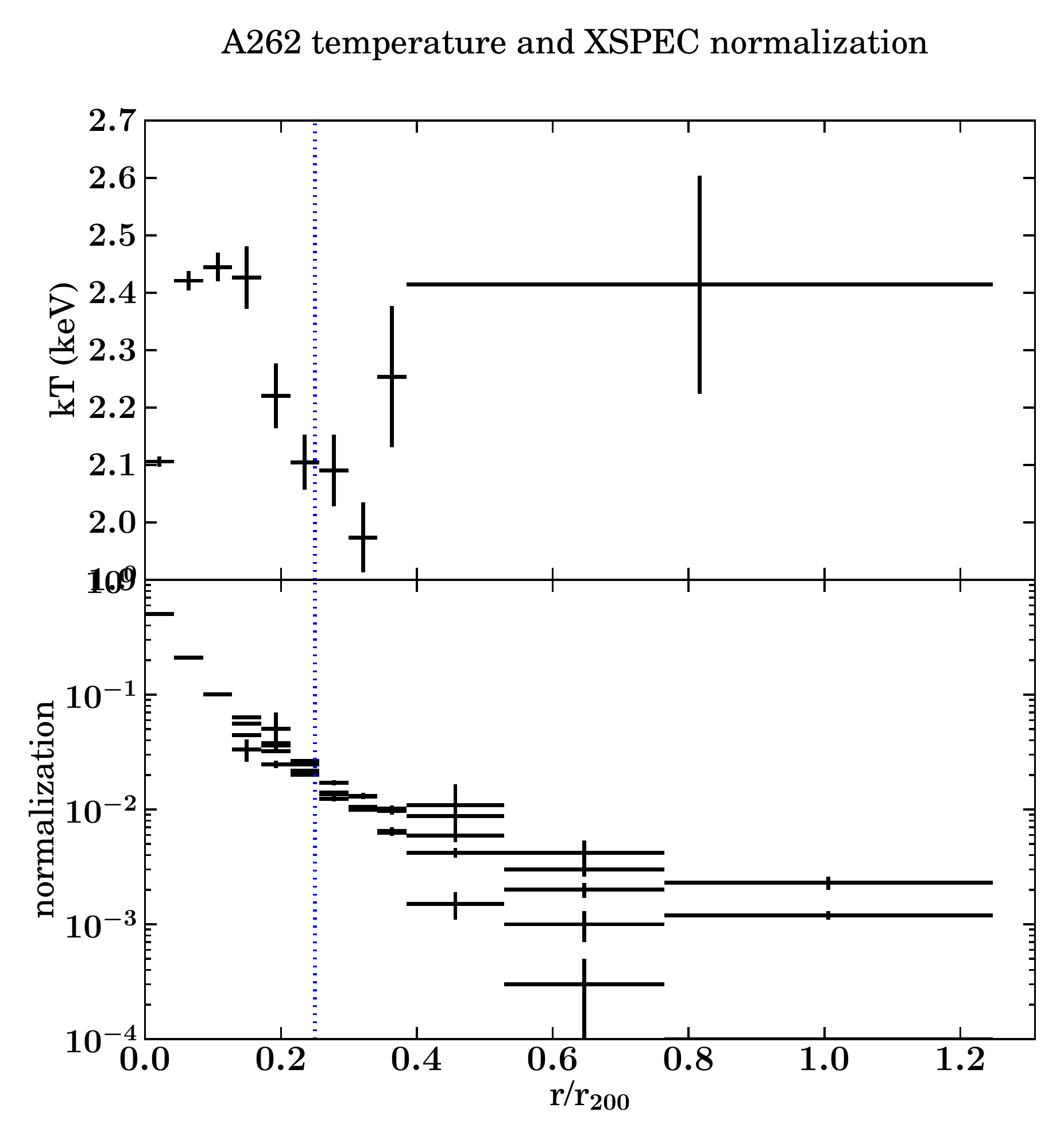} 
\end{minipage}
\begin{minipage}{.47\textwidth}
\includegraphics[width=\textwidth]{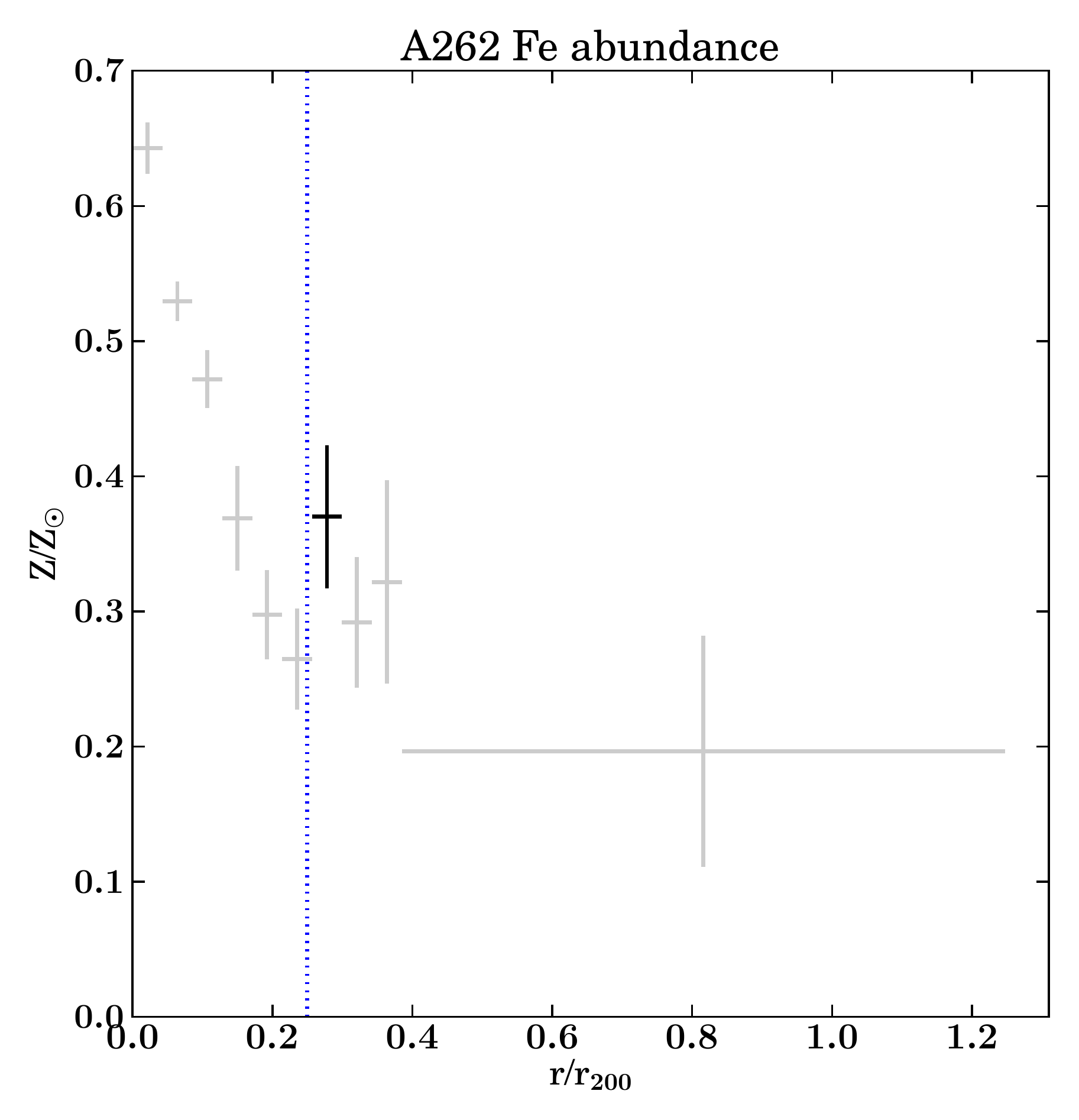}
\end{minipage} 
\caption{\emph{Top left:} The exposure- and vignetting corrected mosaic of
the \suzaku\ observations of~A~262 in the $0.7-7.0$~keV energy range.  The
image has been smoothed with a Gaussian with width of~25~arcsec.  The dashed white circle has a radius of $r_{200}$.  Small magenta circles
show the point sources removed from the spectral analysis. The annular
regions, within which we measured the ICM~metallicity, are shown in yellow. 
The colour bar shows the surface brightness in units of
counts~s$^{-1}$~arcmin$^{-2}$.  
\emph{Top right:} Ratio of the number of the ICM counts to the sum of CXFB and the
instrumental background counts in a 1~keV-wide energy band around the Fe-K
line, in the individual observations (\emph{red}) and for the complete
annulus (\emph{blue}).  The dotted line marks the 10\% threshold employed for all subsequent analysis. For observations where the ratio is close to zero, the data point does not appear in the plot.
\emph{Bottom left:} Projected temperature and normalisation profiles of~A262 (the units of the normalisation are defined in Tab.~\ref{tab:metCxfb}).
The horizontal axes shows the distance from the cluster center in the units of $r_{200}$ shown in
Tab.~\ref{tab:metClusters}.  The vertical dotted line marks $0.25r_{200}$,
which we conservatively assume to be the outside border of the central
metallicity peak. 
\emph{Bottom right:} Radial profile of the best fit projected iron abundance. Only the black data points were included in our subsequent analysis. }
\label{fig:a262portrait} 
\end{figure*}

\begin{figure*}
\begin{minipage}{.49\textwidth}
\includegraphics[width=\textwidth]{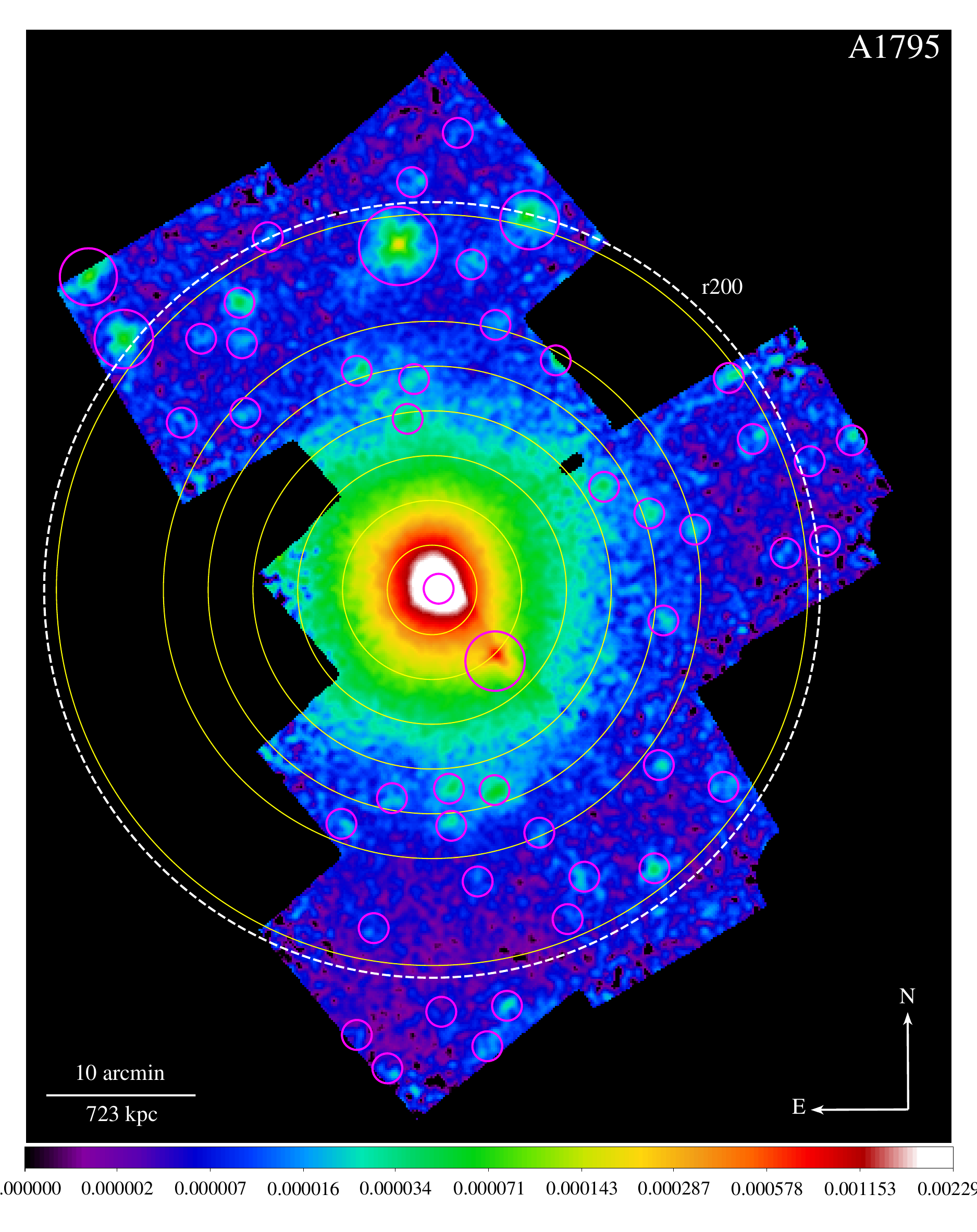}
\end{minipage}
\begin{minipage}{.49\textwidth}
\includegraphics[width=\textwidth]{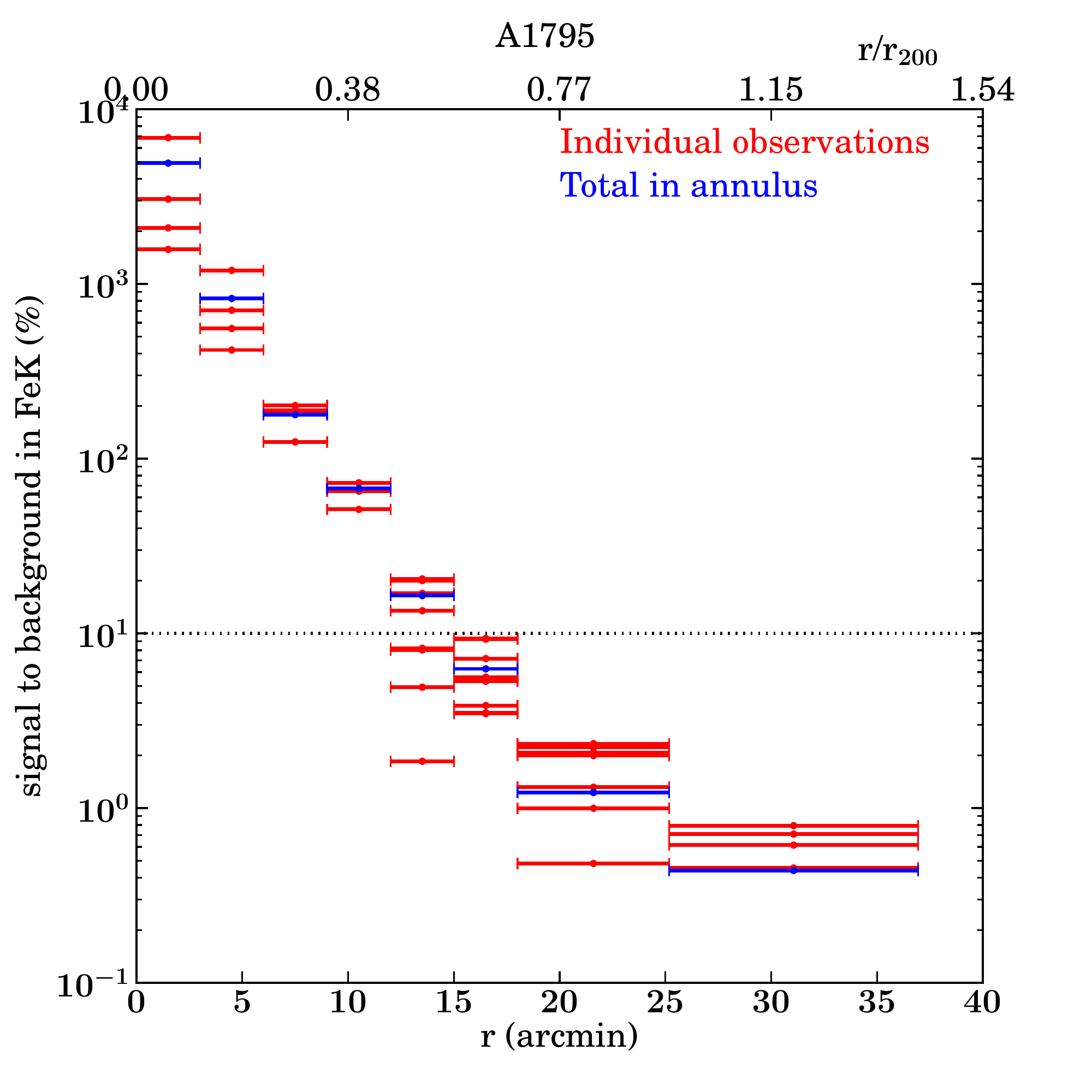}
\end{minipage}
\begin{minipage}{.49\textwidth}
\includegraphics[width=\textwidth]{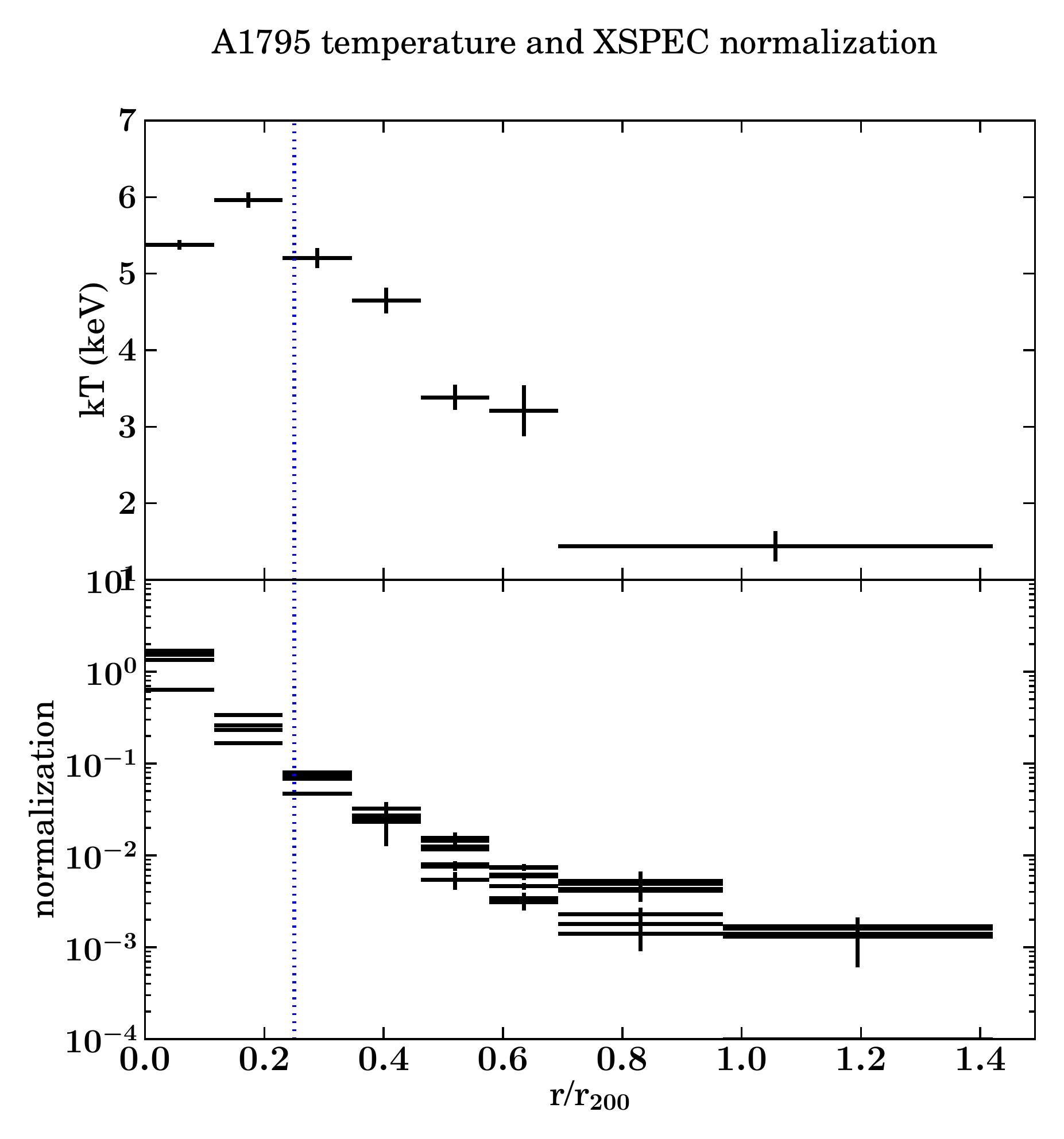}
\end{minipage}
\begin{minipage}{.49\textwidth}
\includegraphics[width=\textwidth]{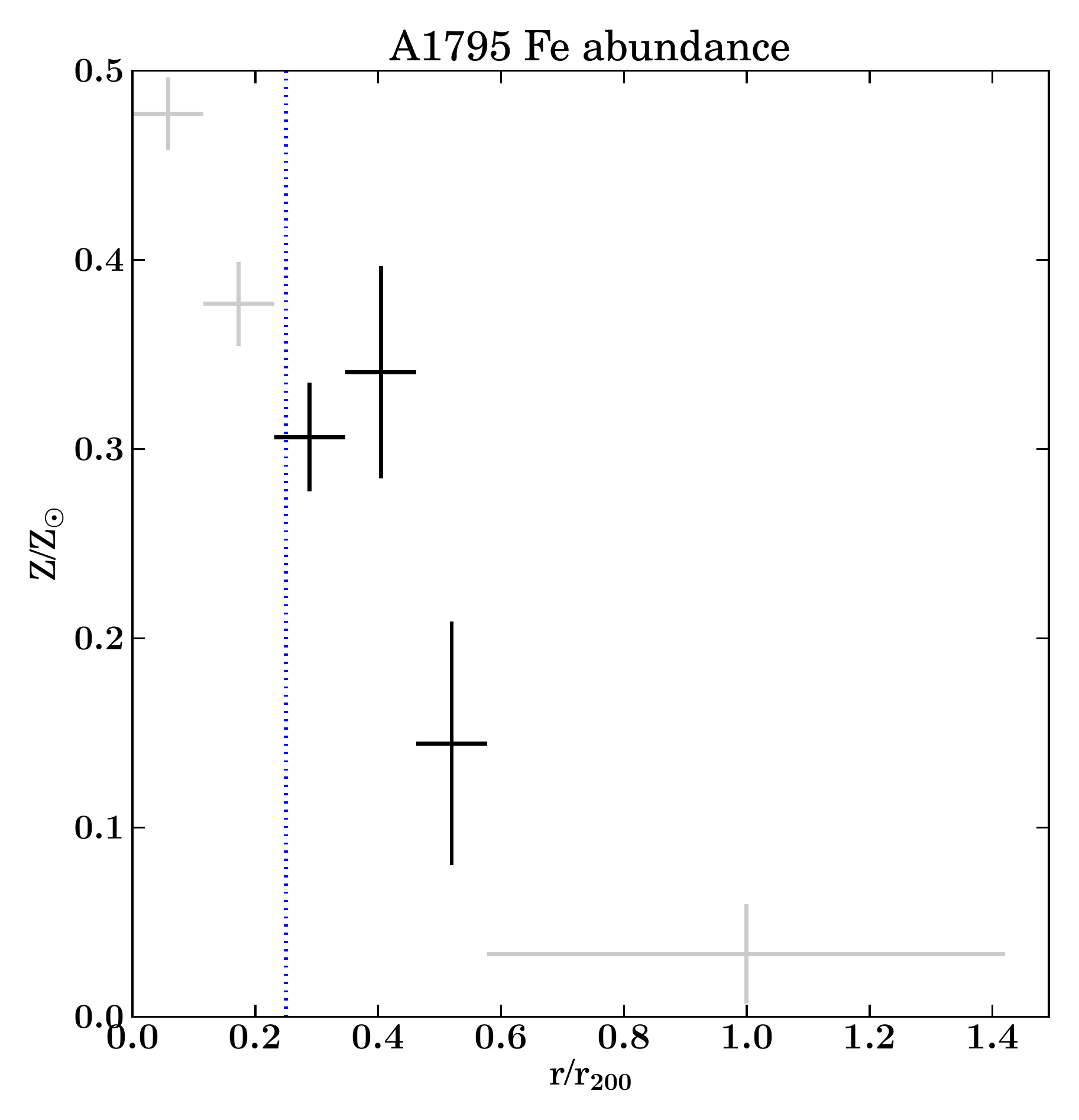}
\end{minipage}
\caption{Same as Fig.~\ref{fig:a262portrait}, but for A~1795.}
\label{fig:a1795portrait}
\end{figure*}

\begin{figure*}
\begin{minipage}{.49\textwidth}
\includegraphics[width=\textwidth]{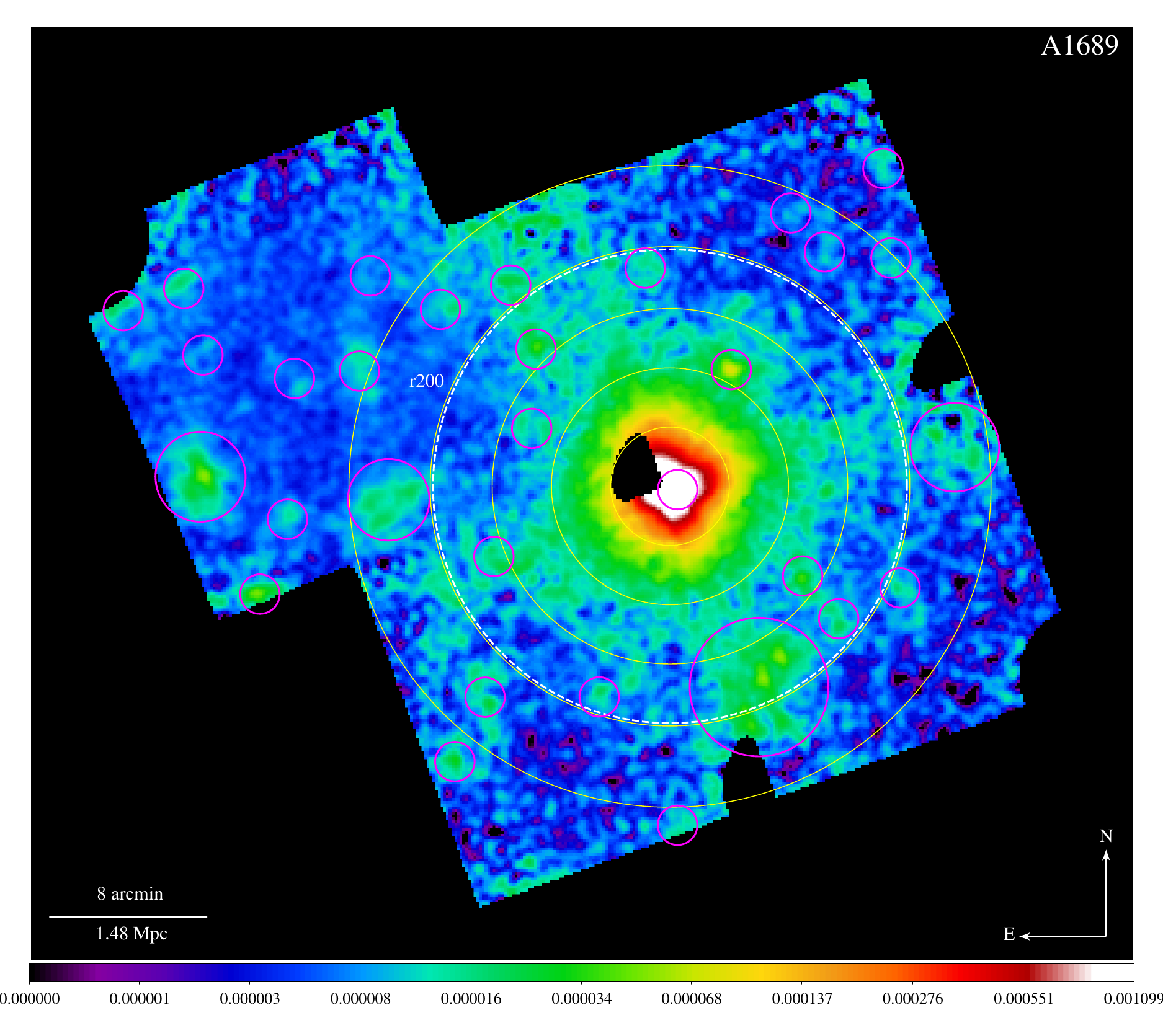}
\end{minipage}
\begin{minipage}{.49\textwidth}
\includegraphics[width=\textwidth]{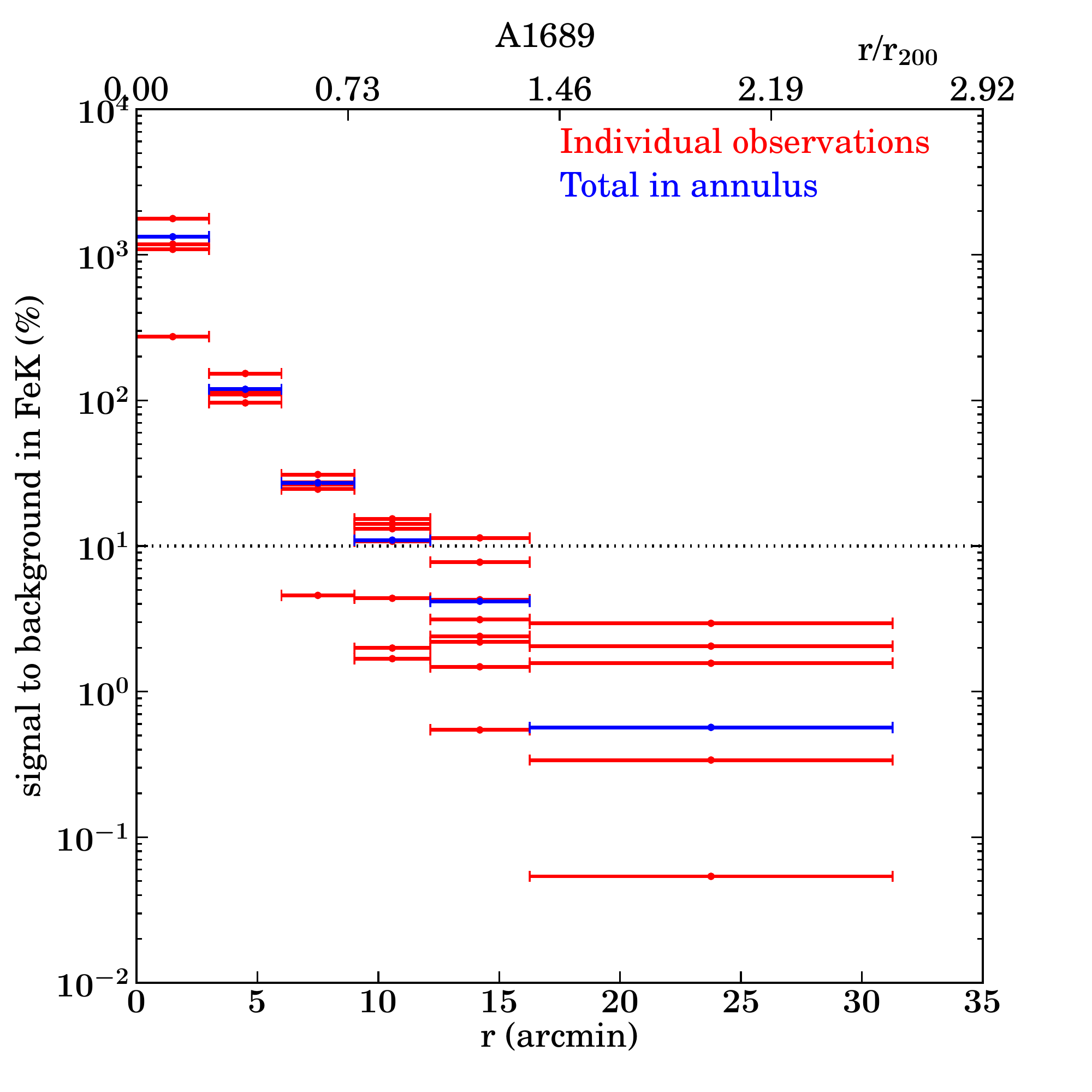}
\end{minipage}
\begin{minipage}{.49\textwidth}
\includegraphics[width=\textwidth]{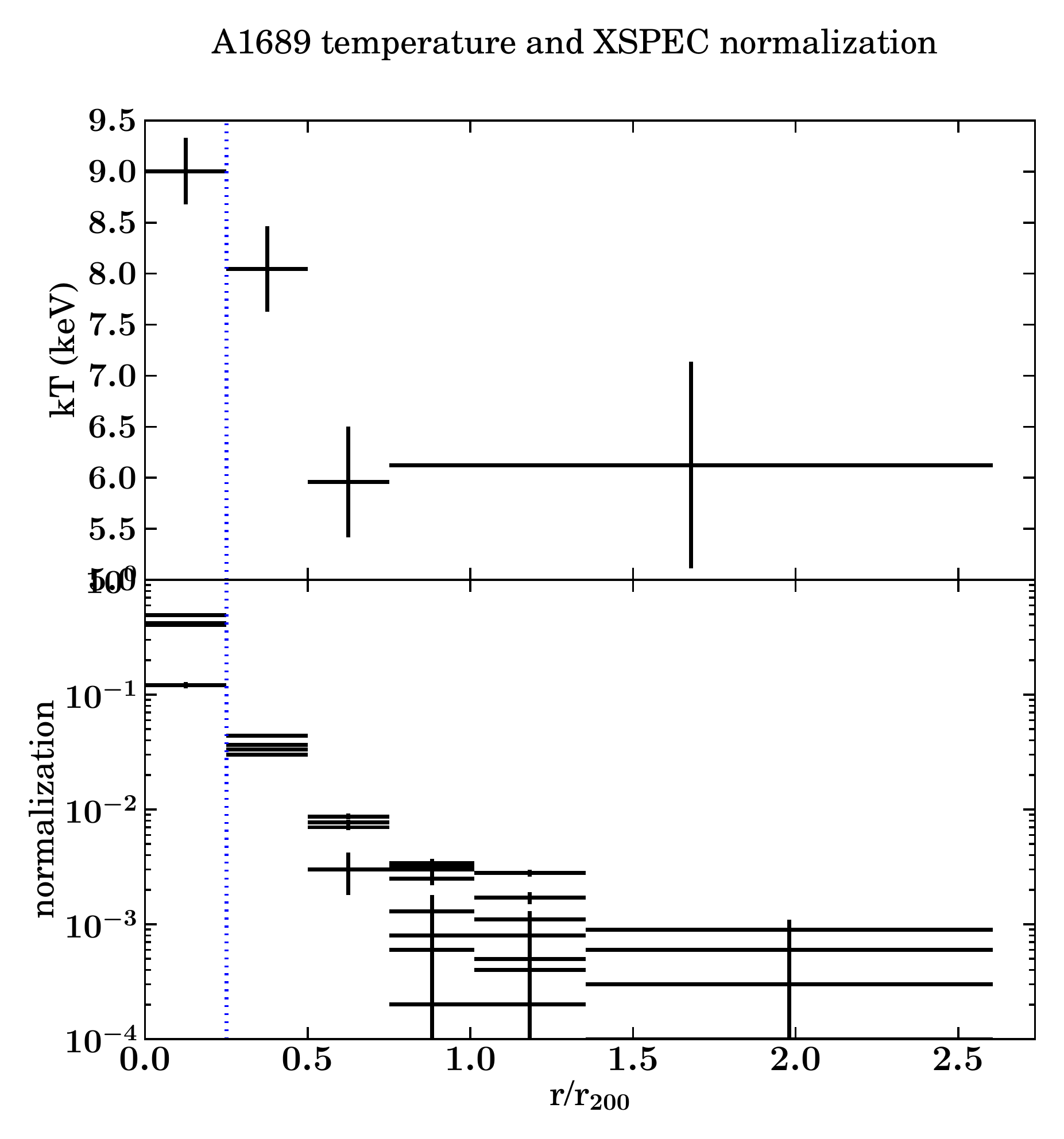}
\end{minipage}
\begin{minipage}{.49\textwidth}
\includegraphics[width=\textwidth]{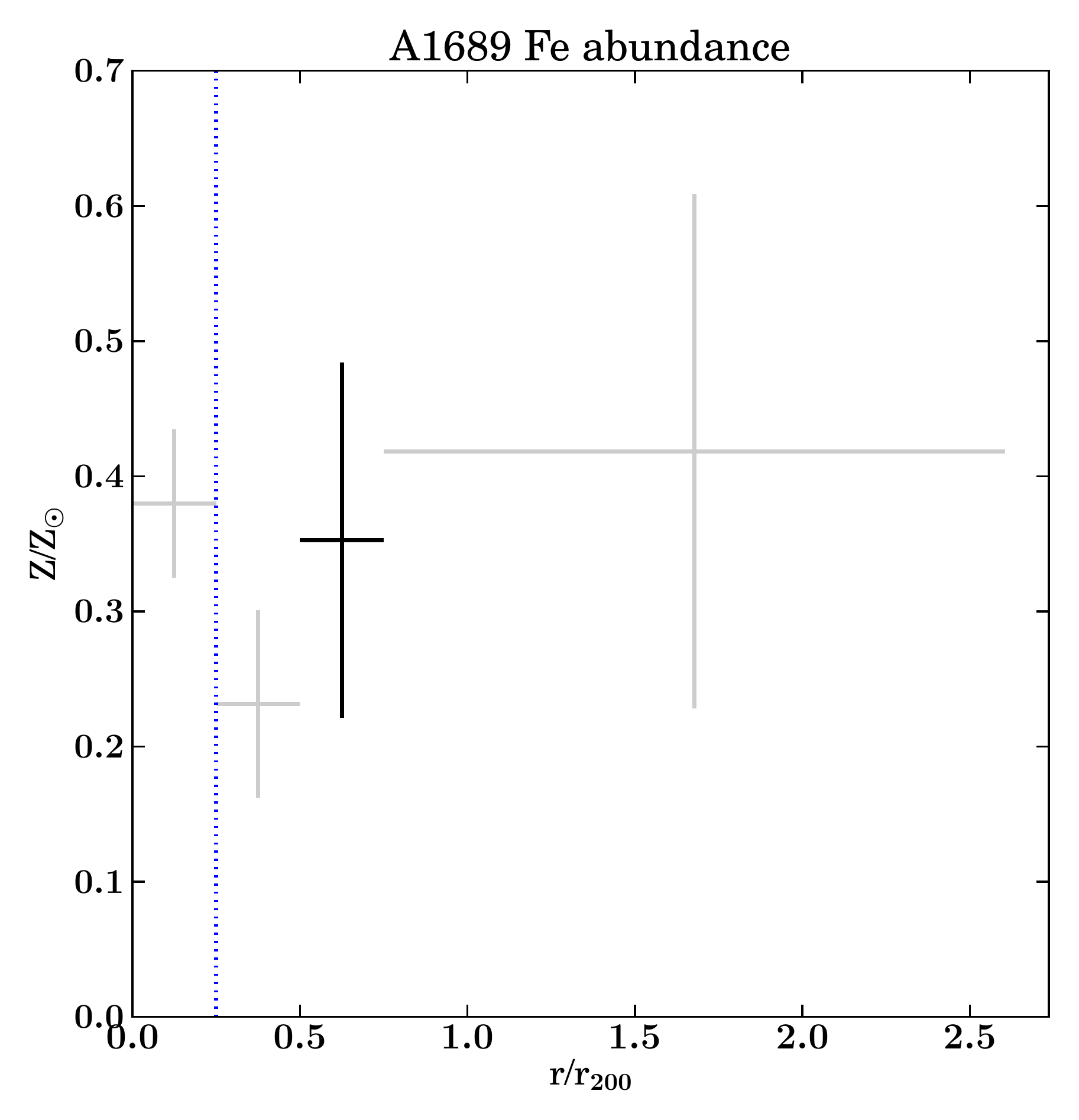}
\end{minipage}
\caption{Same as Fig.~\ref{fig:a262portrait}, but for A~1689.}
\label{fig:a1689portrait}
\end{figure*}

\begin{figure*}
\begin{minipage}{.49\textwidth}
\includegraphics[width=\textwidth]{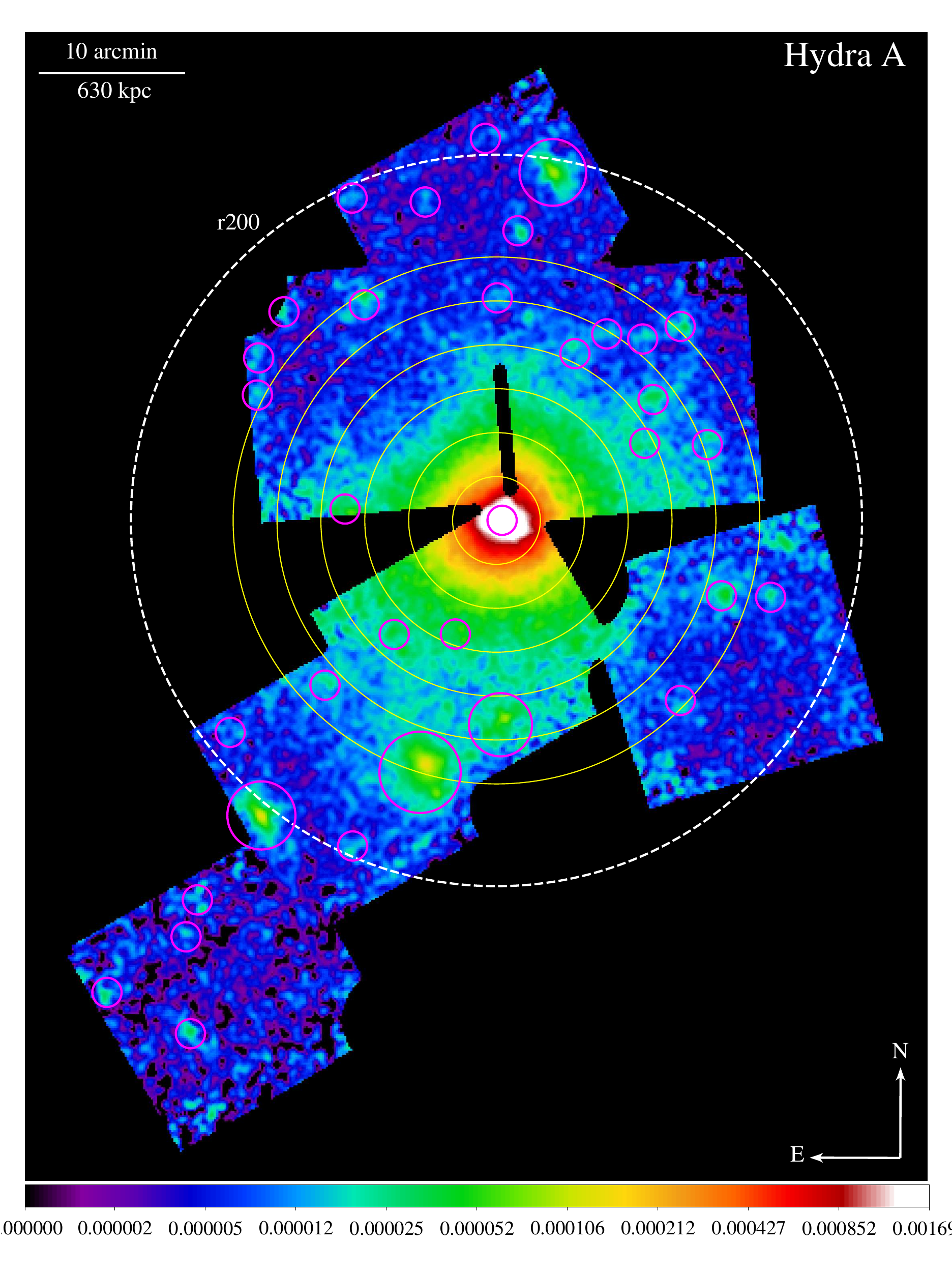}
\end{minipage}
\begin{minipage}{.49\textwidth}
\includegraphics[width=\textwidth]{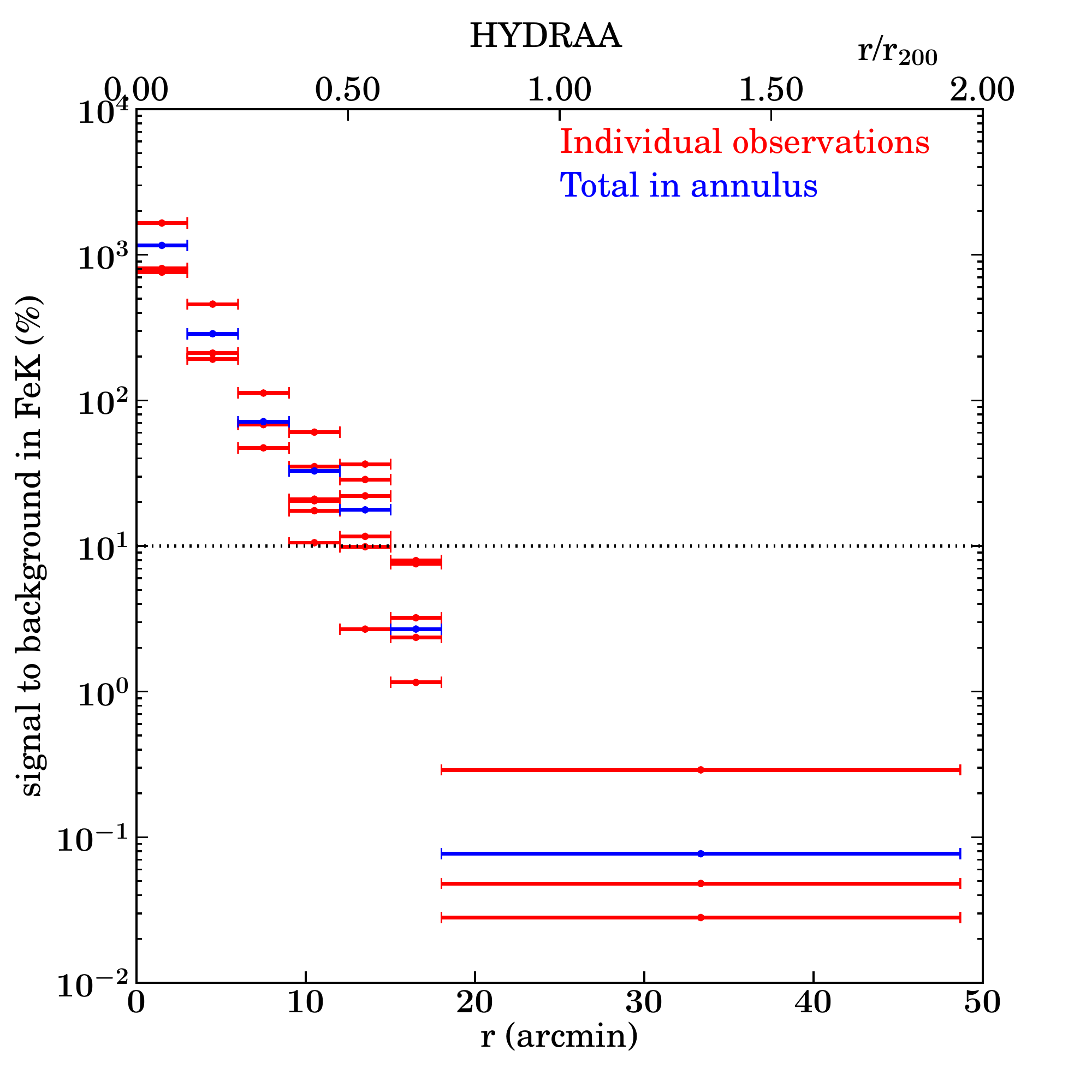}
\end{minipage}
\begin{minipage}{.49\textwidth}
\includegraphics[width=\textwidth]{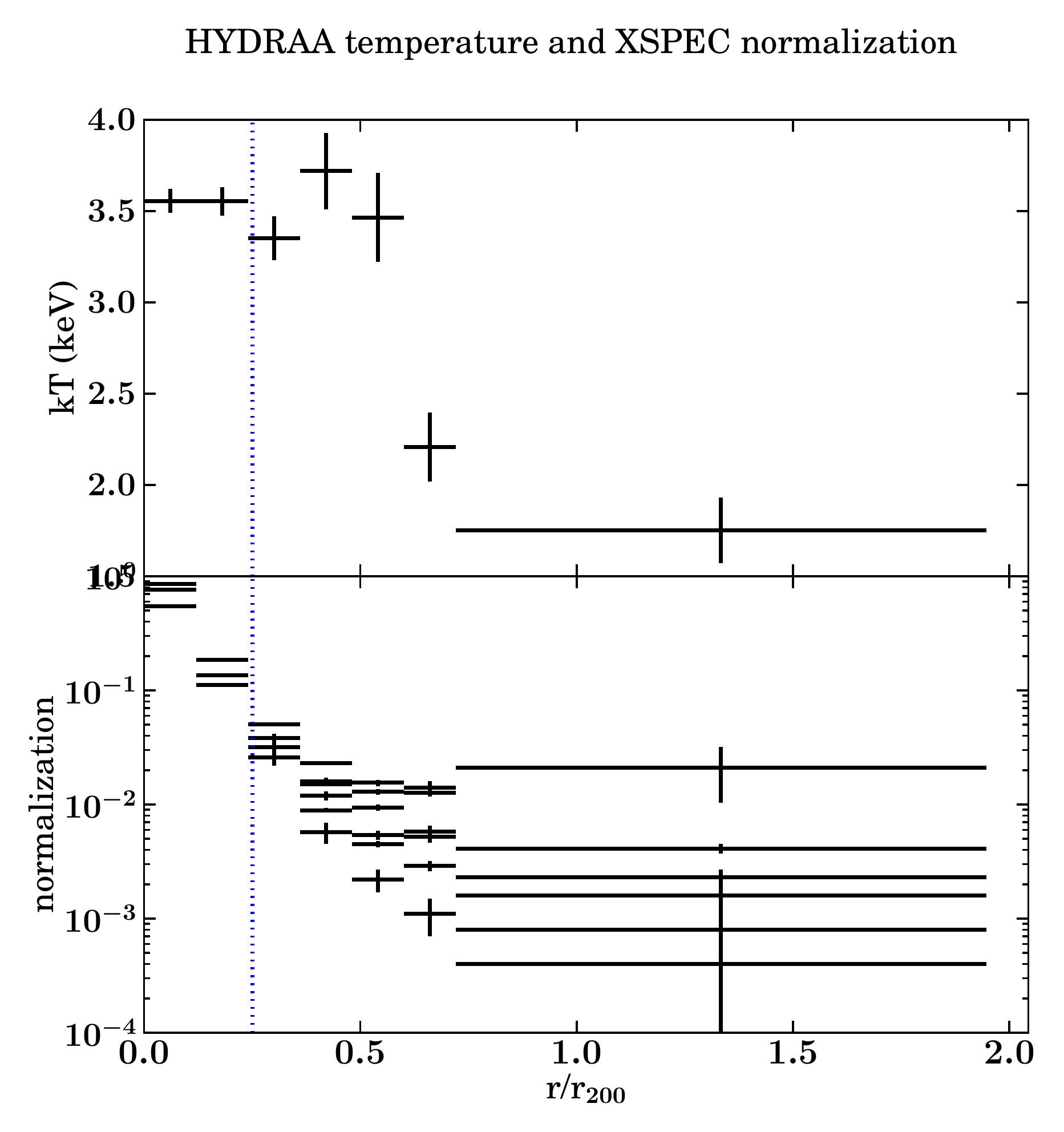}
\end{minipage}
\begin{minipage}{.49\textwidth}
\includegraphics[width=\textwidth]{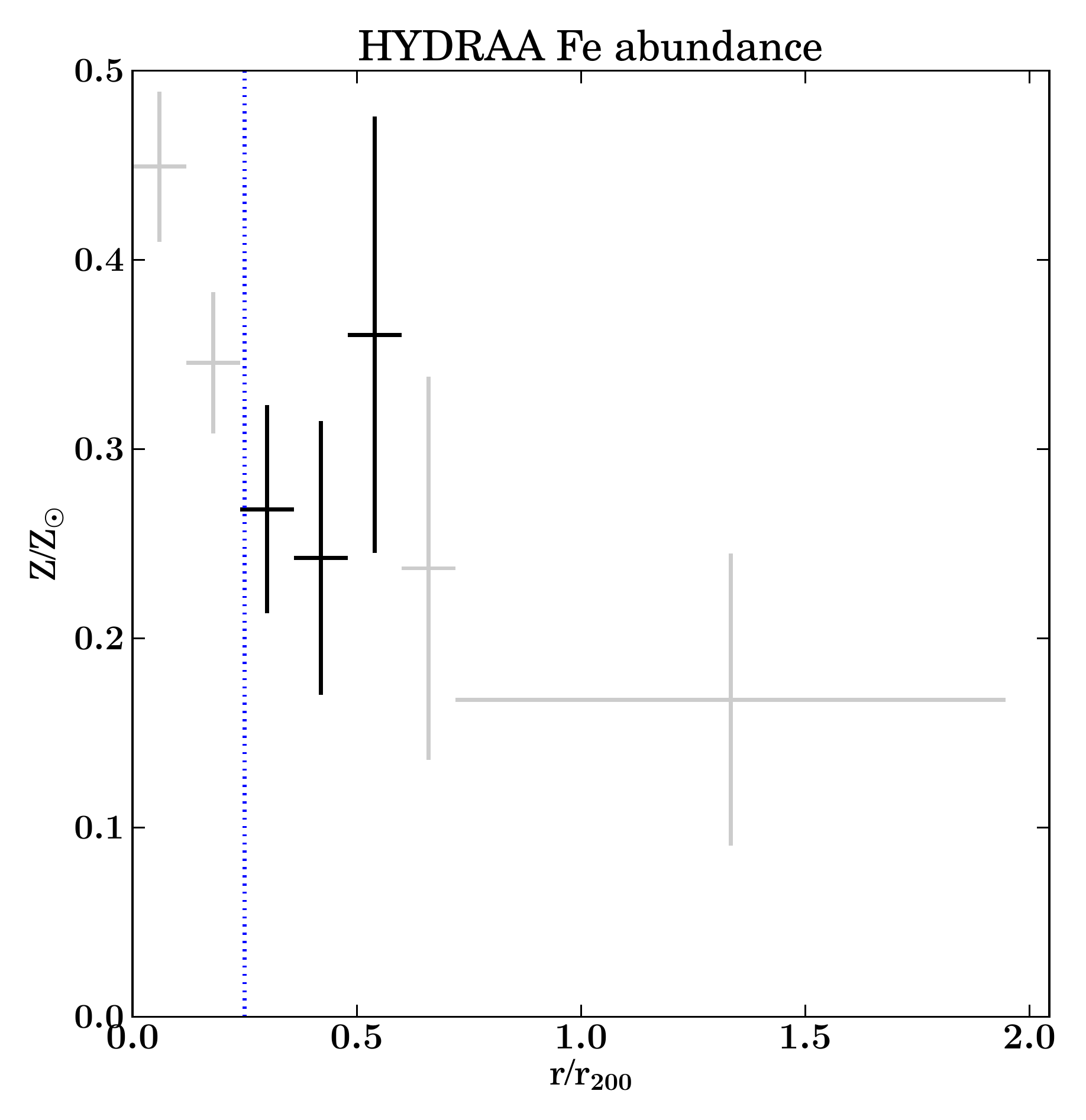}
\end{minipage}
\caption{Same as Fig.~\ref{fig:a262portrait}, but for Hydra~A.}
\label{fig:hydraAportrait}
\end{figure*}

\begin{figure*}
\begin{minipage}{.49\textwidth}
\includegraphics[width=\textwidth]{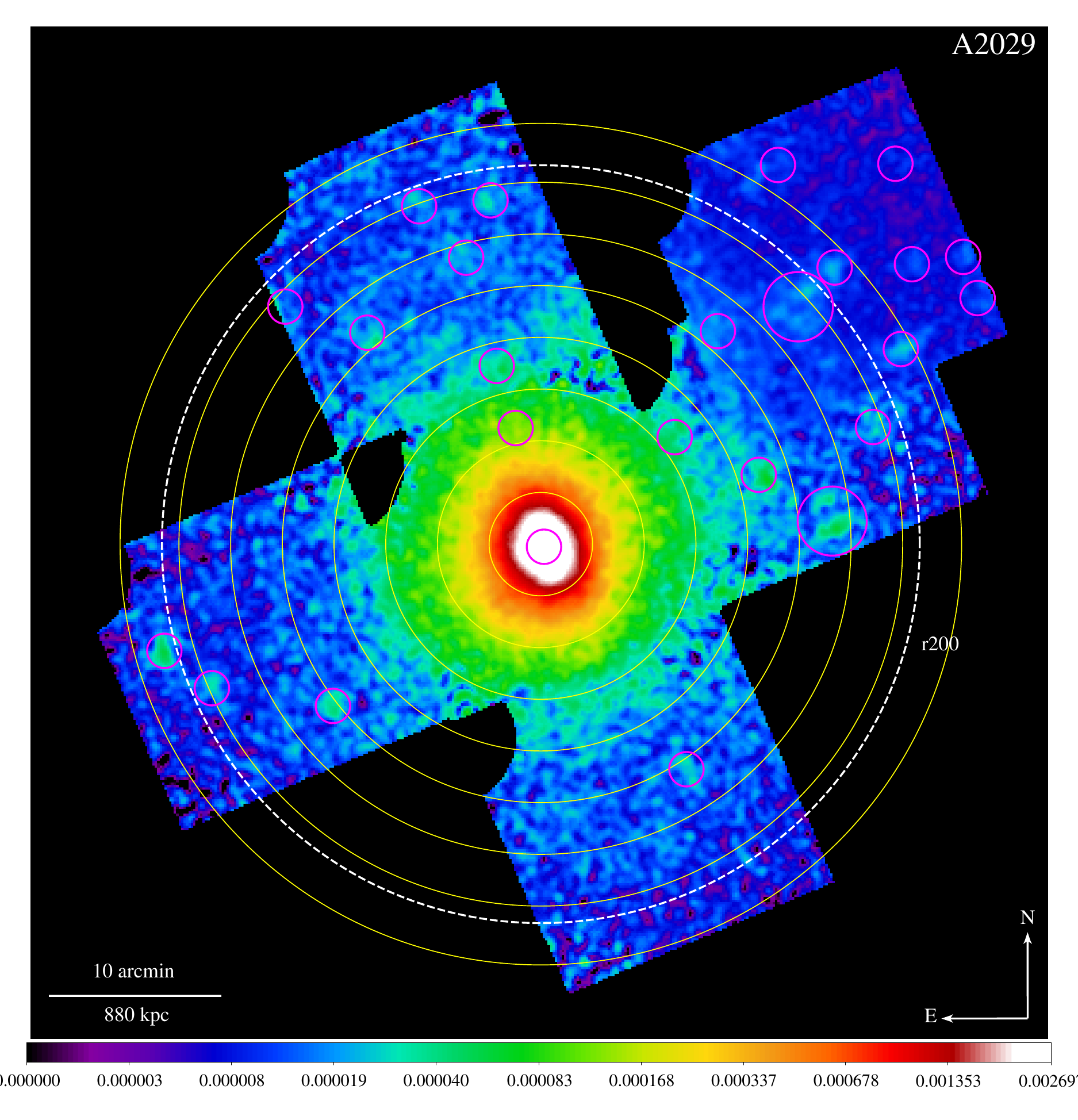}
\end{minipage}
\begin{minipage}{.49\textwidth}
\includegraphics[width=\textwidth]{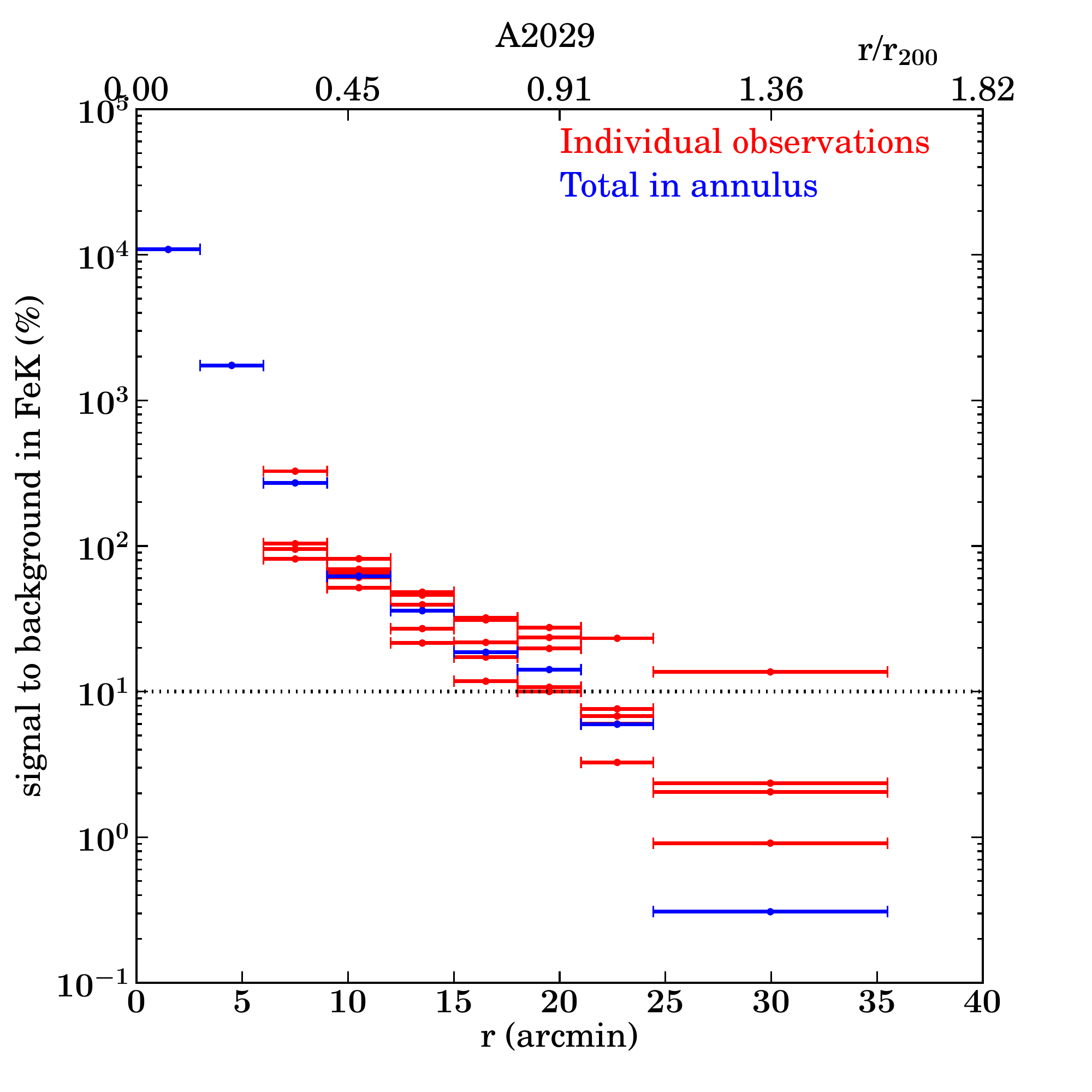}
\end{minipage}
\begin{minipage}{.49\textwidth}
\includegraphics[width=\textwidth]{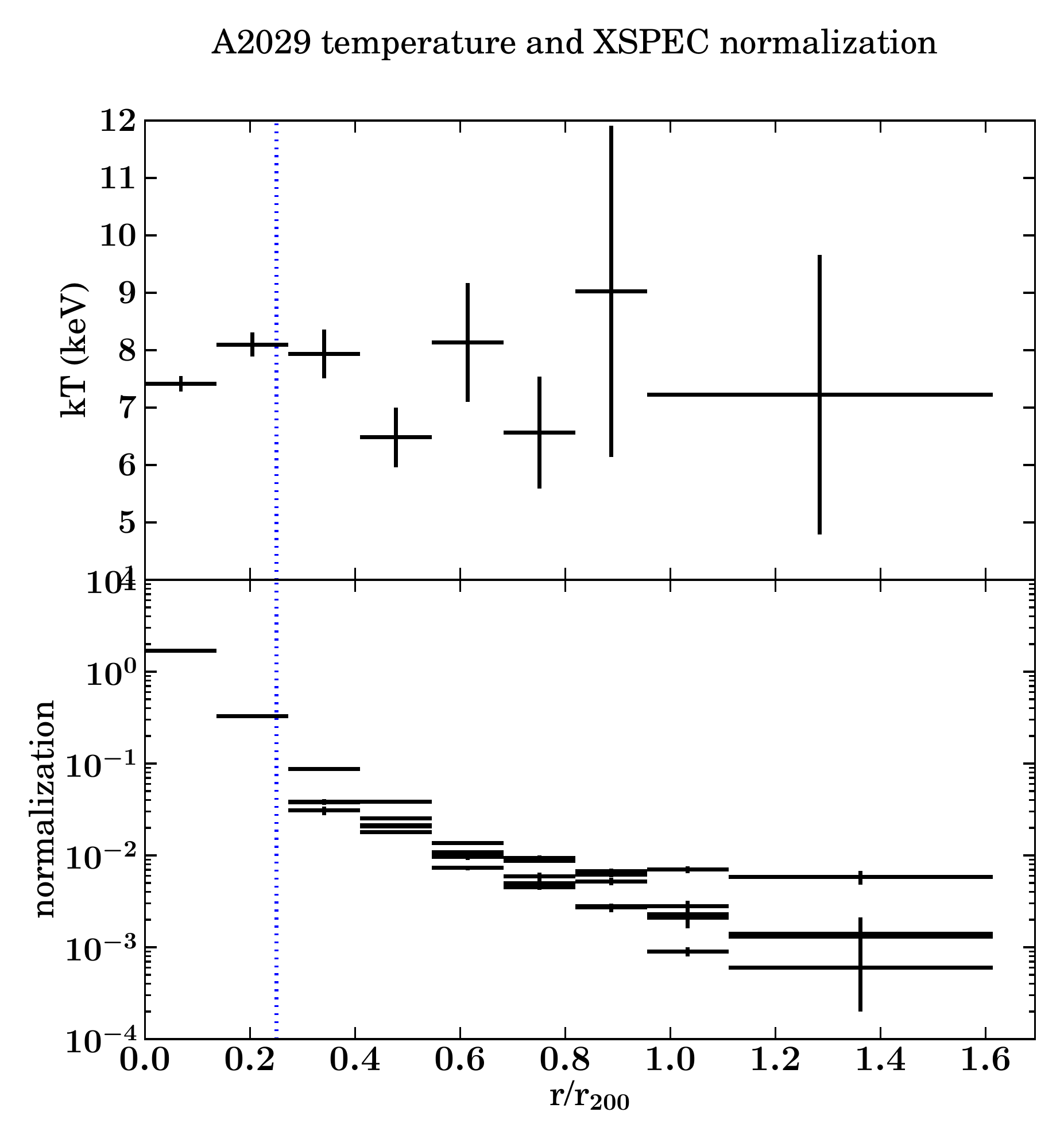}
\end{minipage}
\begin{minipage}{.49\textwidth}
\includegraphics[width=\textwidth]{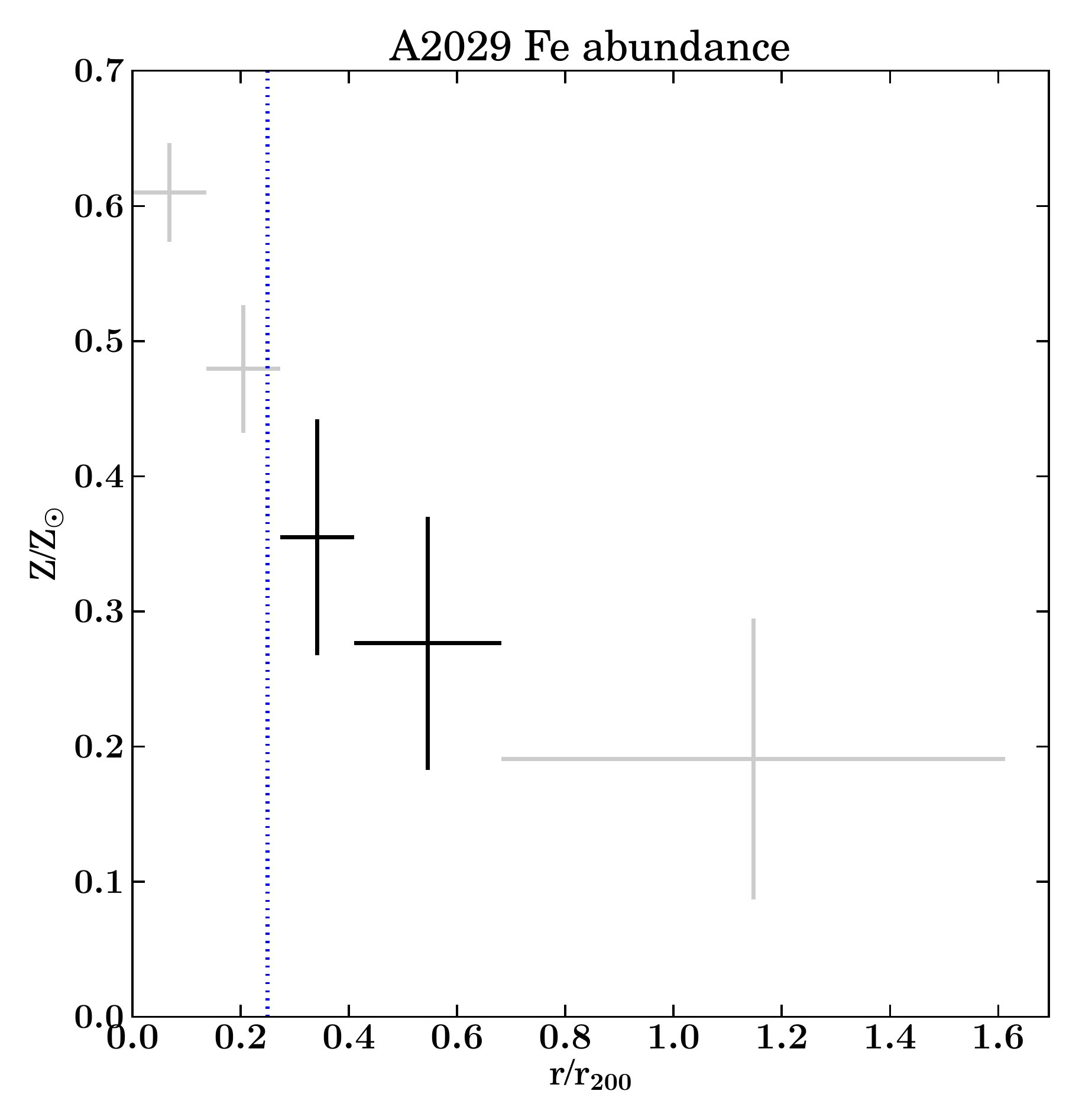}
\end{minipage}
\caption{Same as Fig.~\ref{fig:a262portrait}, but for A~2029.}
\label{fig:a2029portrait}
\end{figure*}

\begin{figure*}
\begin{minipage}{.49\textwidth}
\includegraphics[width=\textwidth]{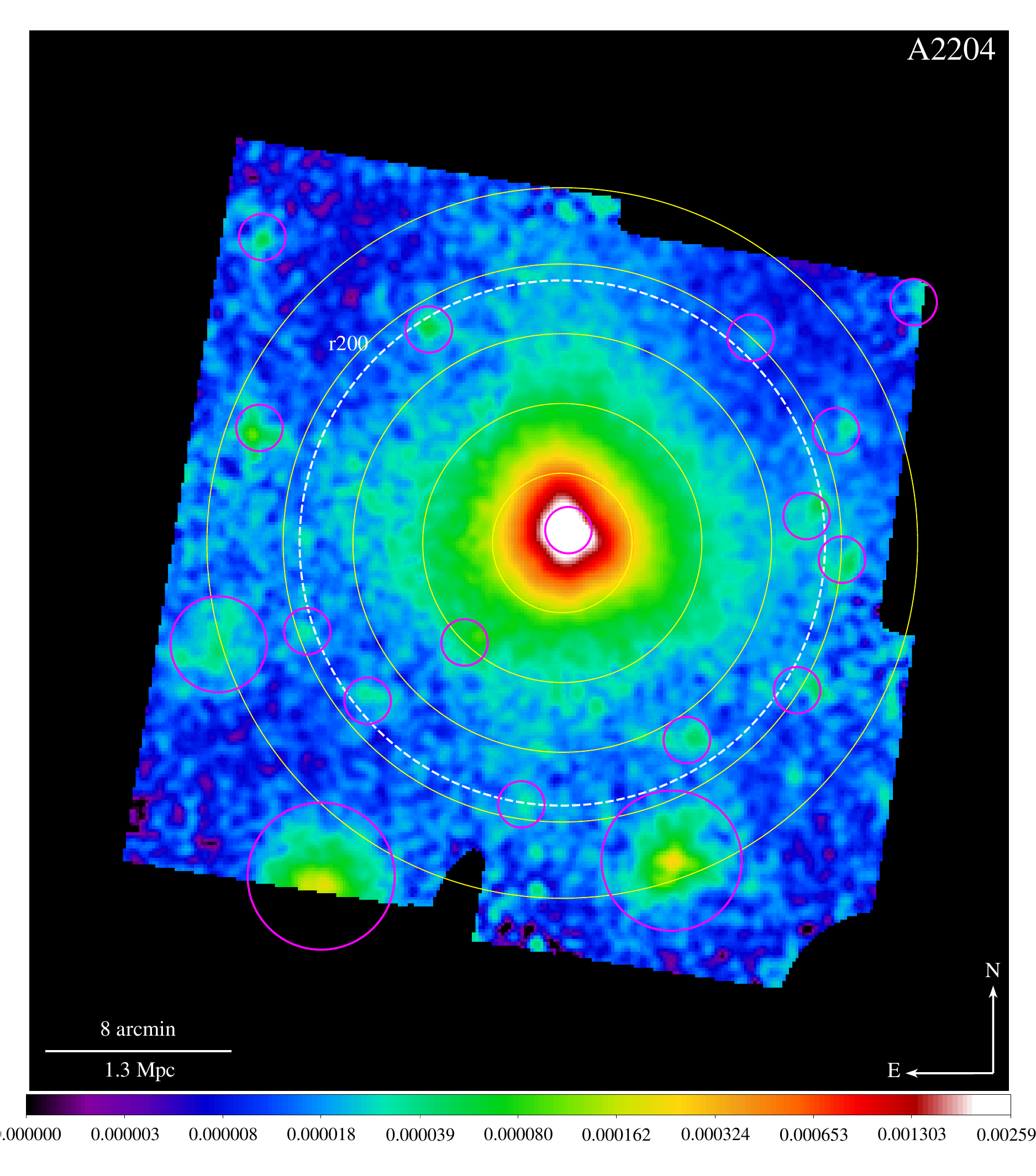}
\end{minipage}
\begin{minipage}{.49\textwidth}
\includegraphics[width=\textwidth]{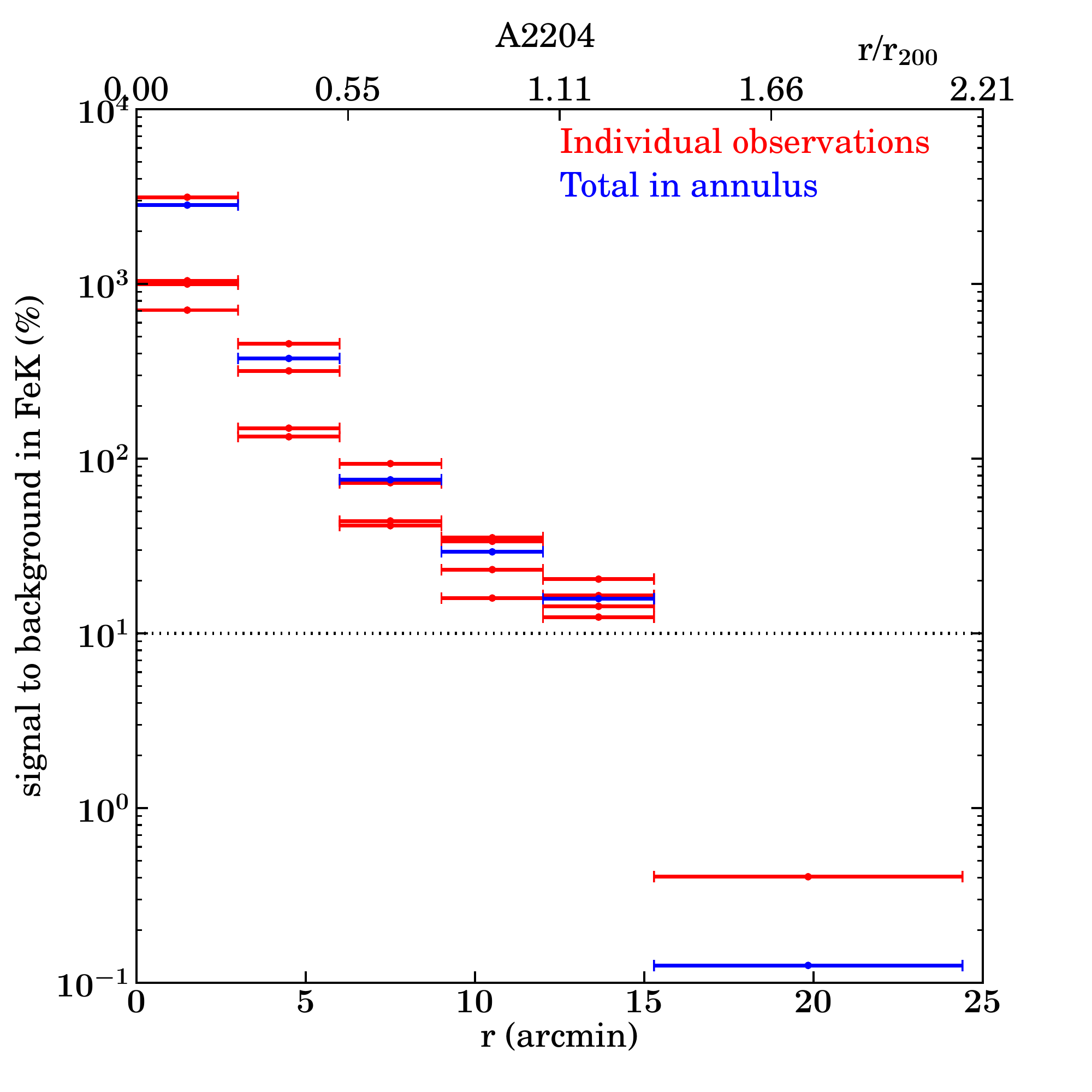}
\end{minipage}
\begin{minipage}{.49\textwidth}
\includegraphics[width=\textwidth]{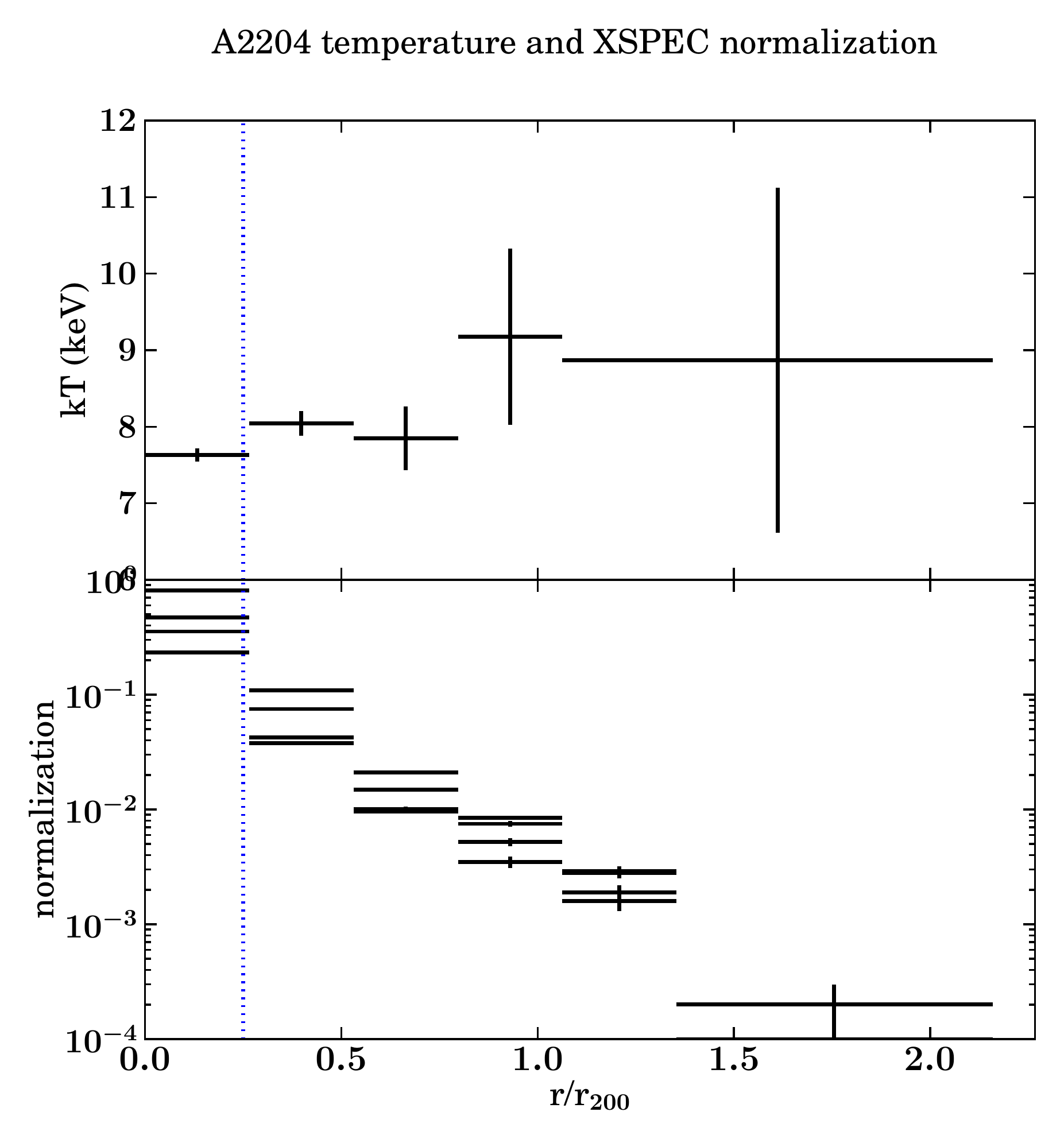}
\end{minipage}
\begin{minipage}{.49\textwidth}
\includegraphics[width=\textwidth]{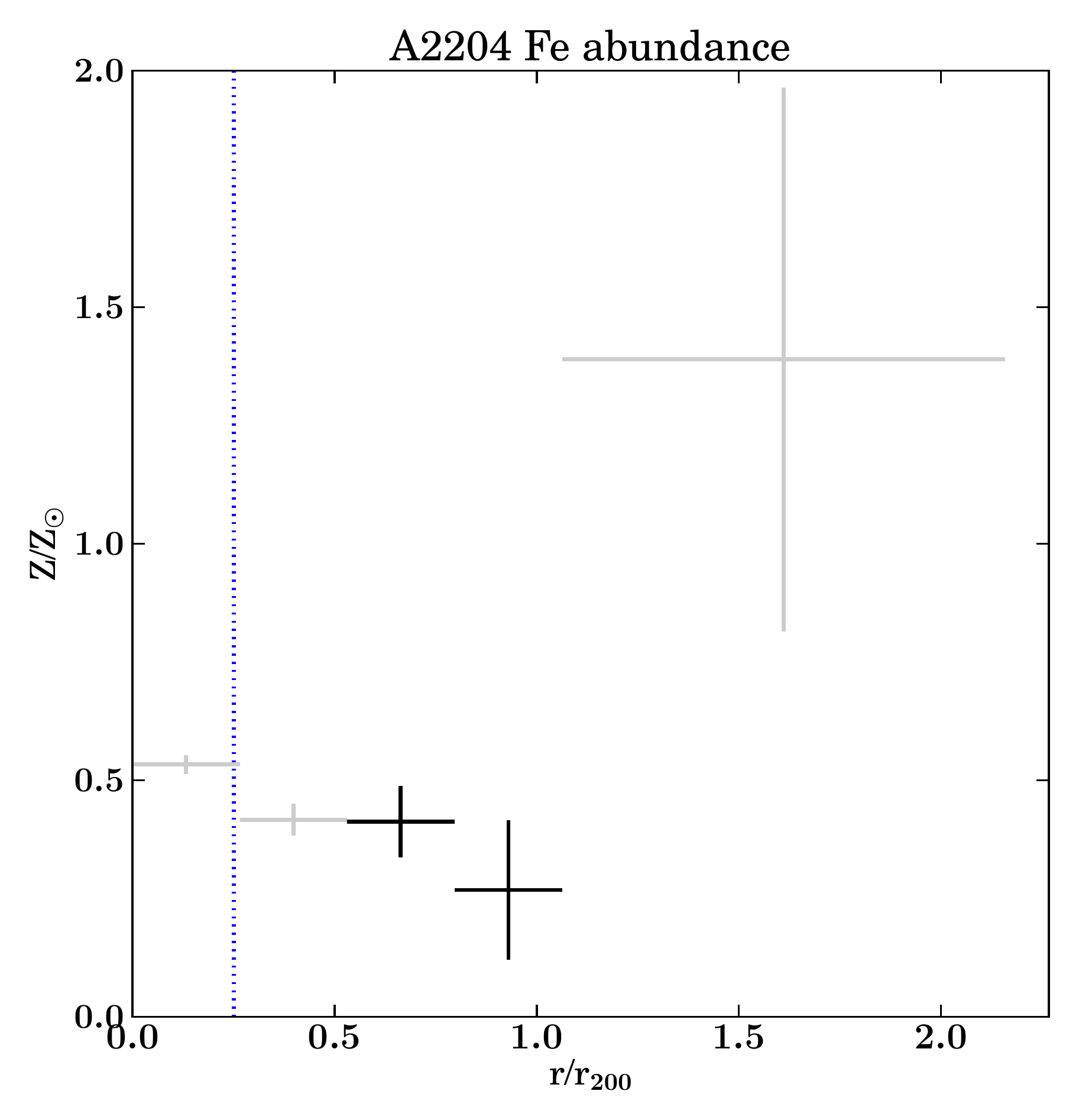}
\end{minipage}
\caption{Same as Fig.~\ref{fig:a262portrait}, but for A~2204.}
\label{fig:a2204portrait}
\end{figure*}

\begin{figure*}
\begin{minipage}{.49\textwidth}
\includegraphics[width=\textwidth]{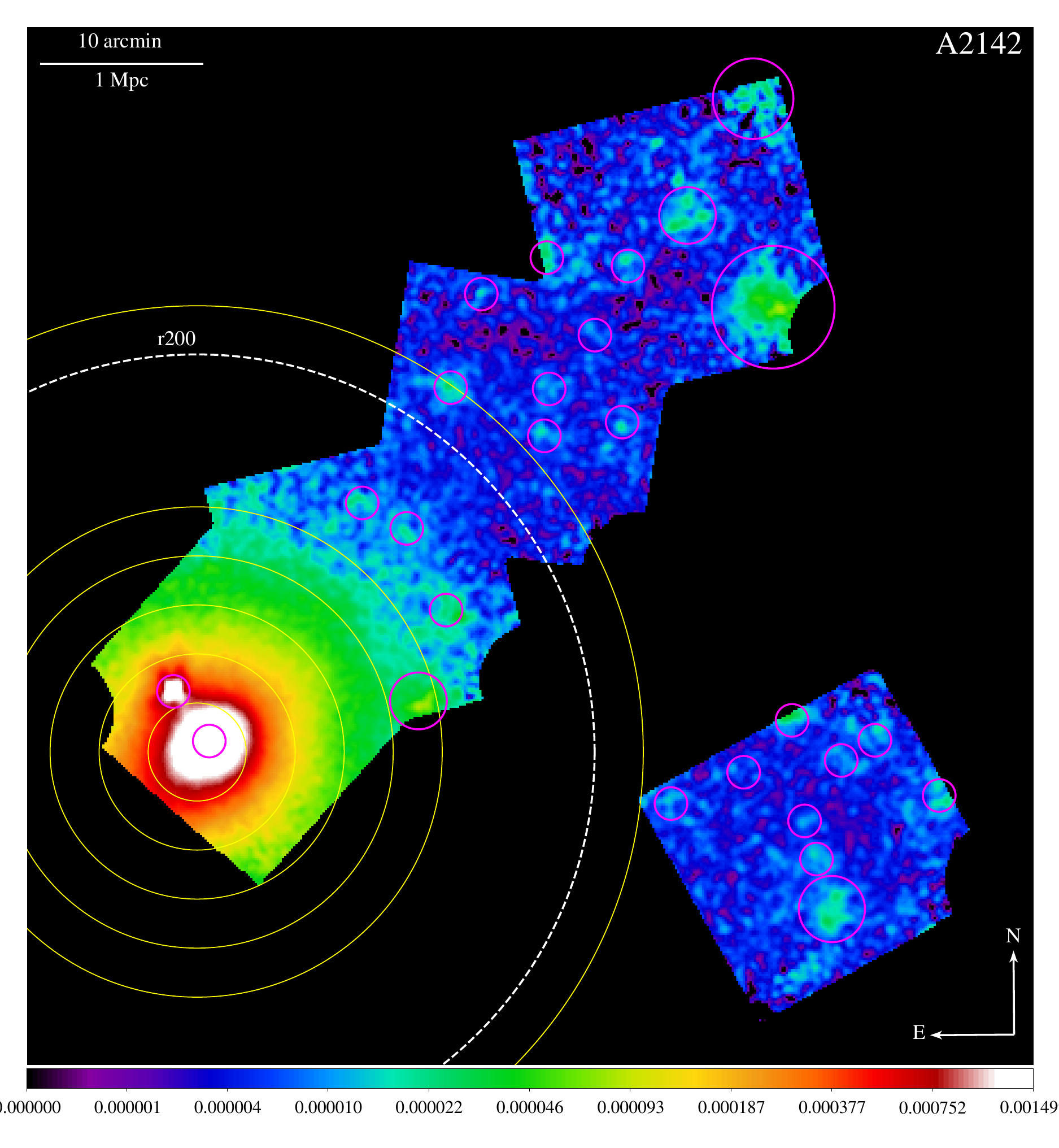}
\end{minipage}
\begin{minipage}{.49\textwidth}
\includegraphics[width=\textwidth]{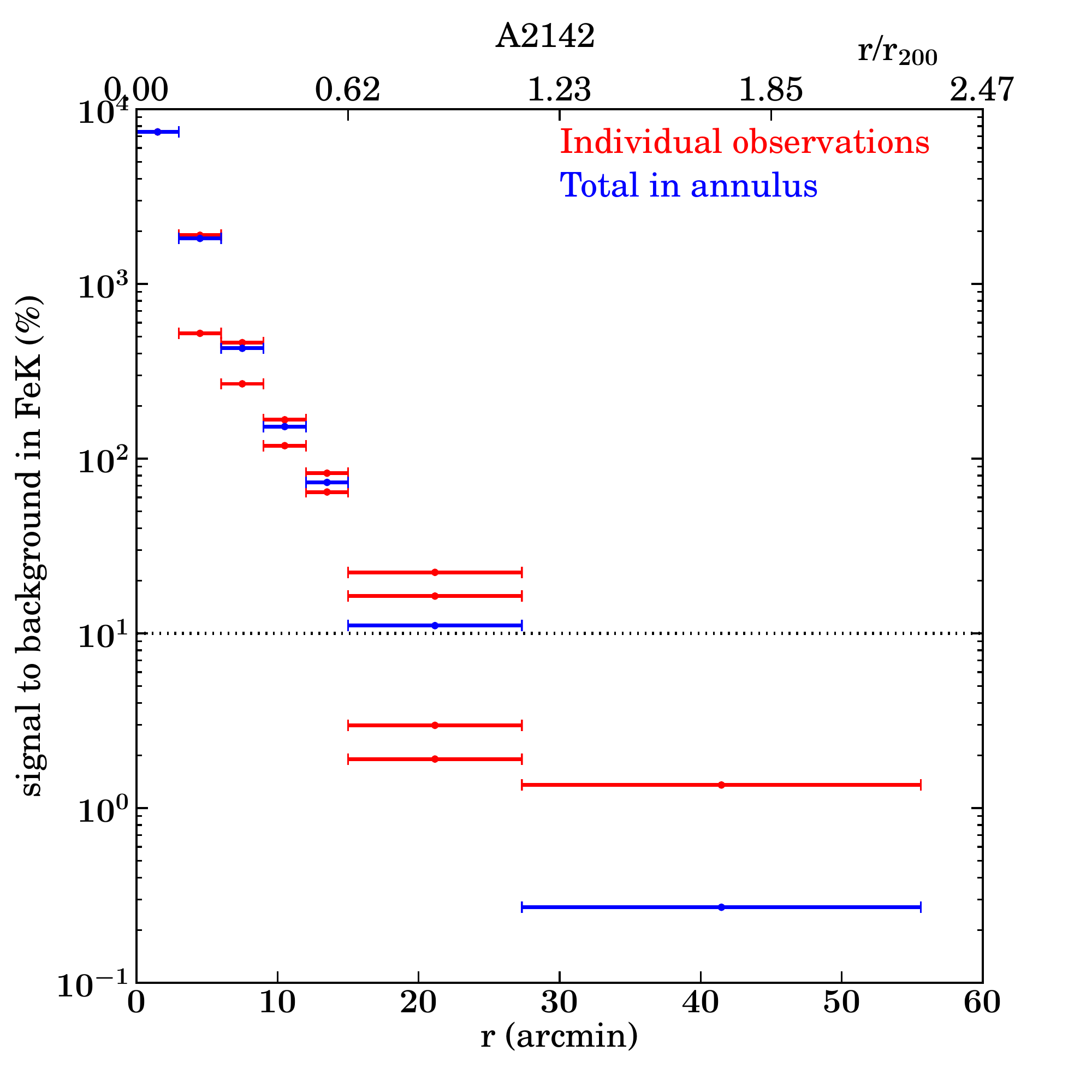}
\end{minipage}
\begin{minipage}{.49\textwidth}
\includegraphics[width=\textwidth]{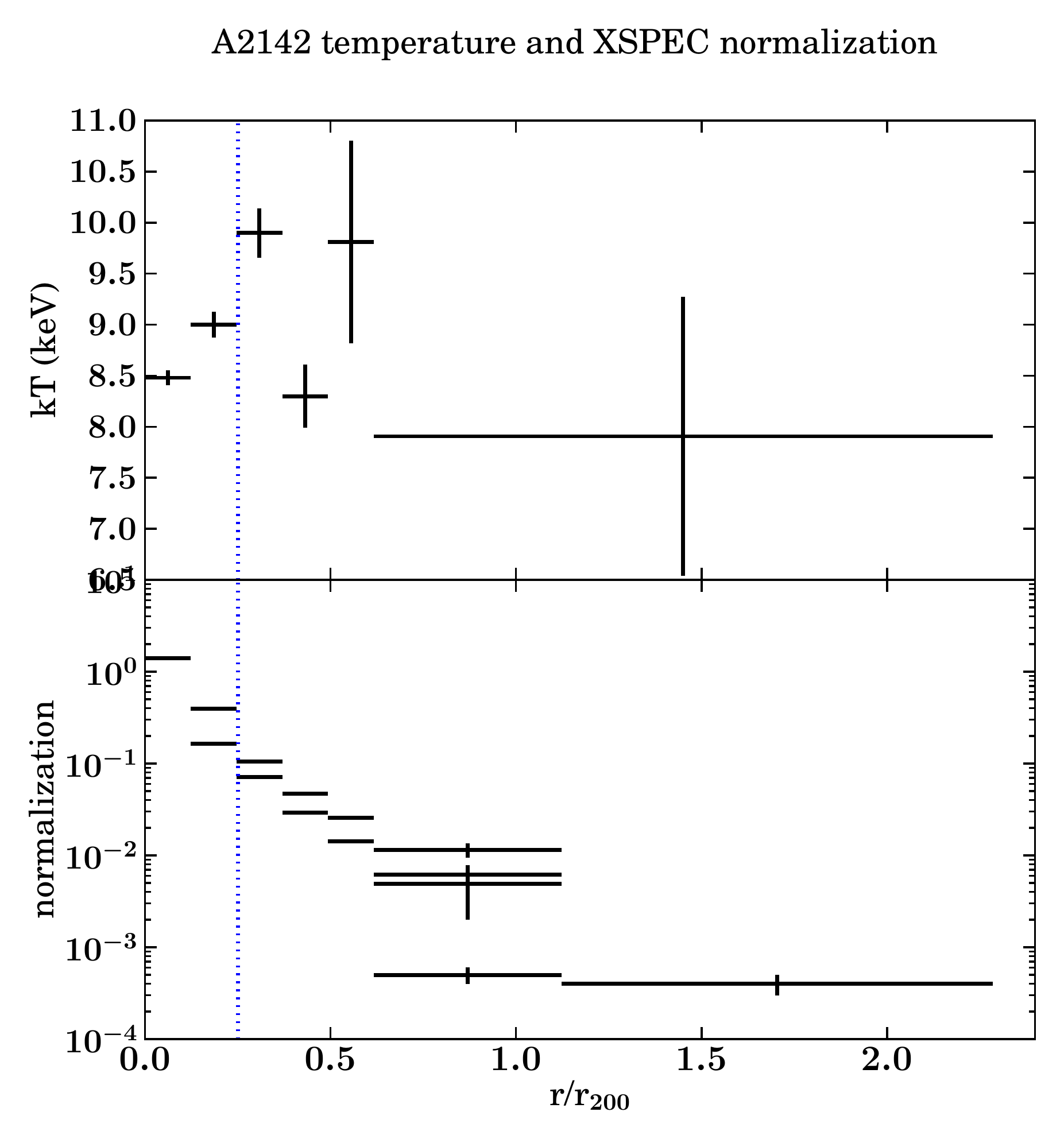}
\end{minipage}
\begin{minipage}{.49\textwidth}
\includegraphics[width=\textwidth]{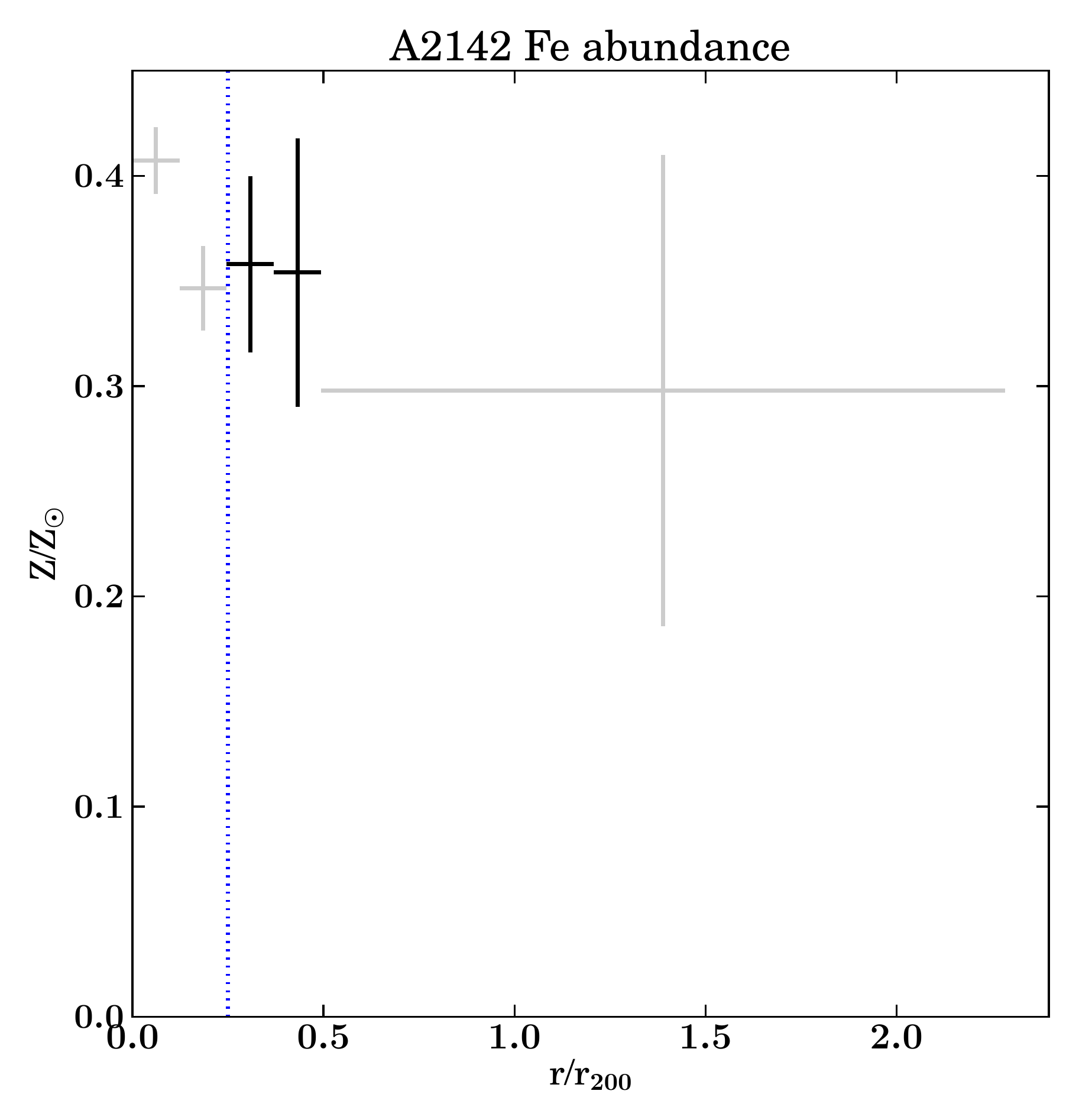}
\end{minipage}
\caption{Same as Fig.~\ref{fig:a262portrait}, but for A~2142.}
\label{fig:a2142portrait}
\end{figure*}

\begin{figure*}
\begin{minipage}{.49\textwidth}
\includegraphics[width=\textwidth]{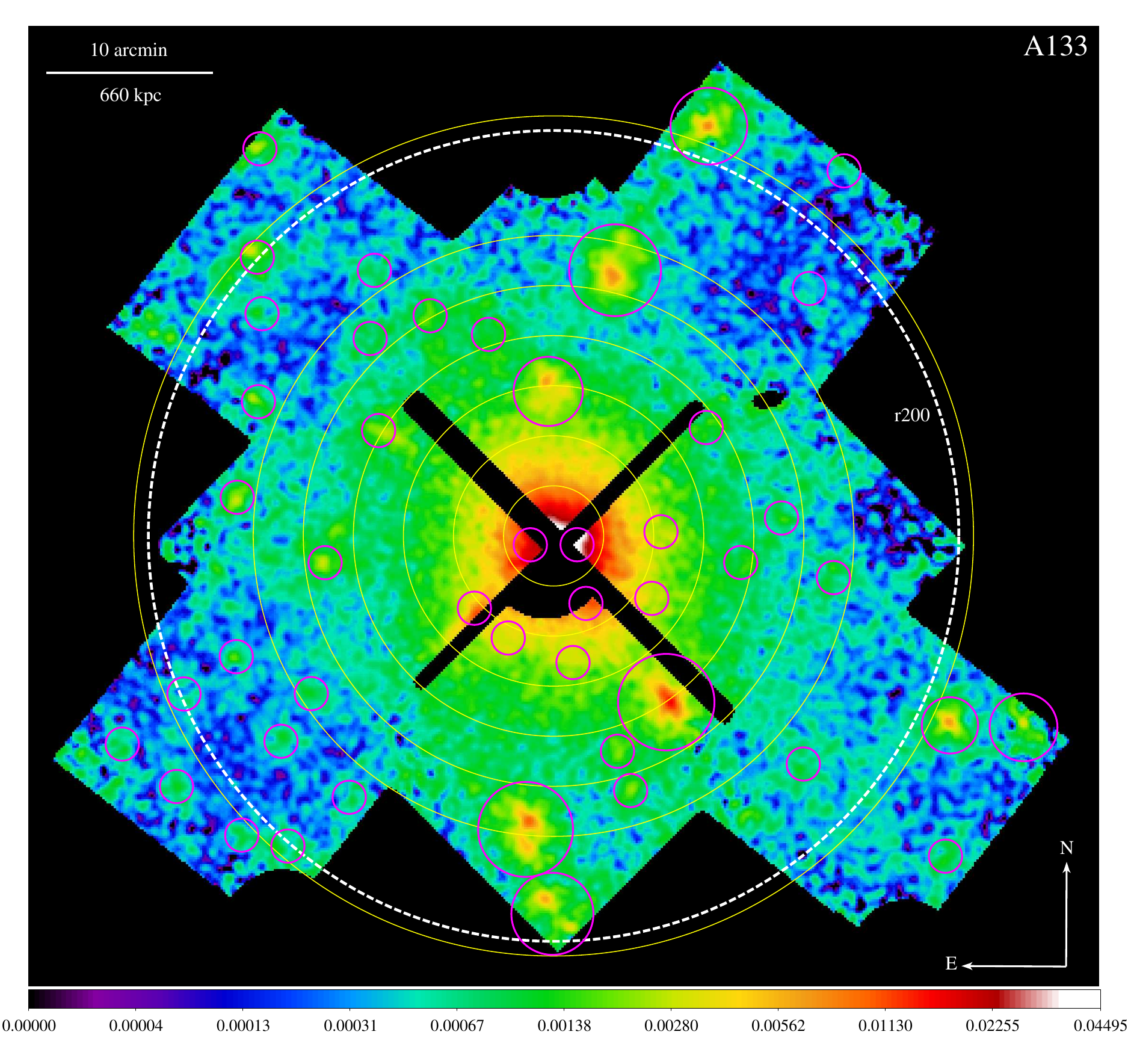}
\end{minipage}
\begin{minipage}{.49\textwidth}
\includegraphics[width=\textwidth]{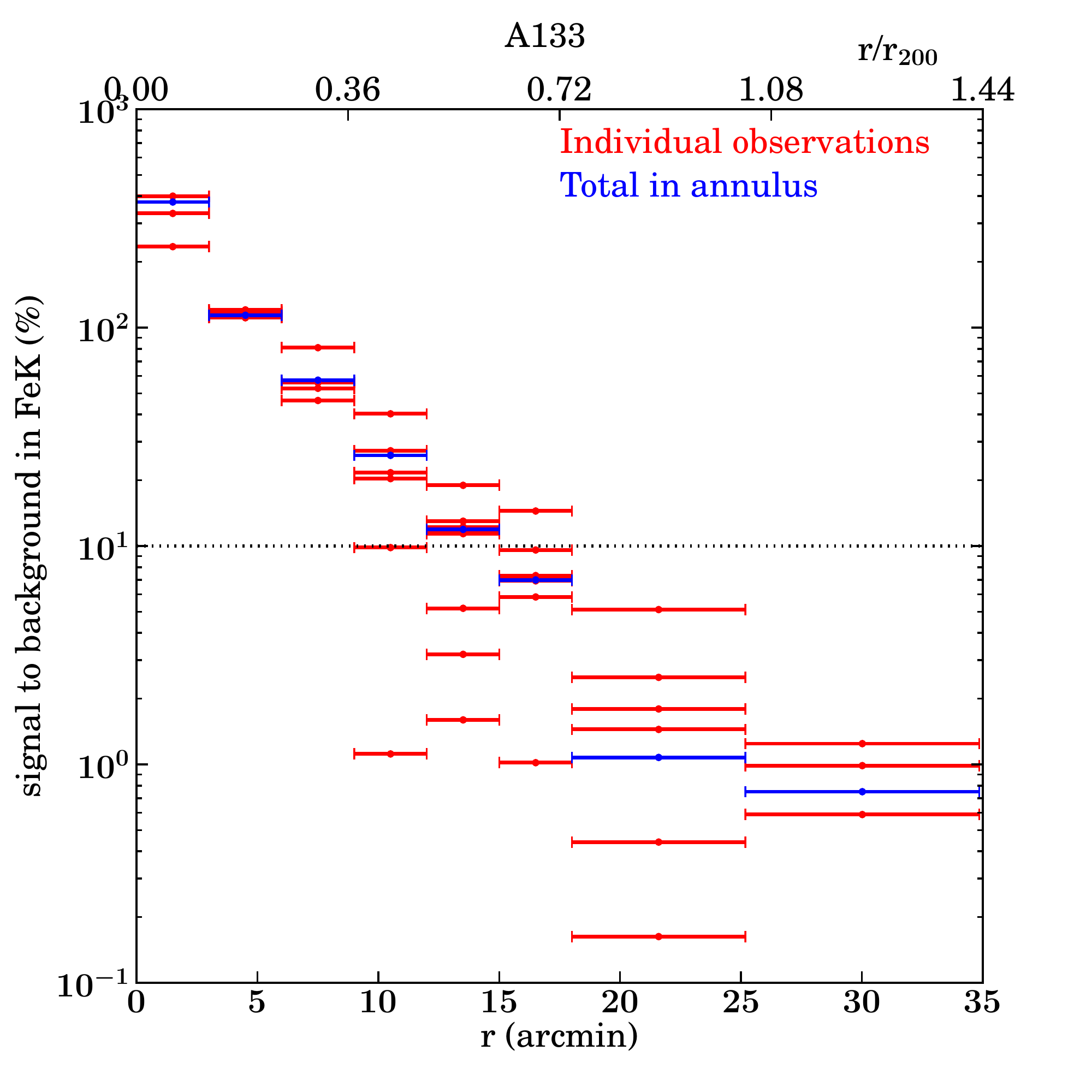}
\end{minipage}
\begin{minipage}{.49\textwidth}
\includegraphics[width=\textwidth]{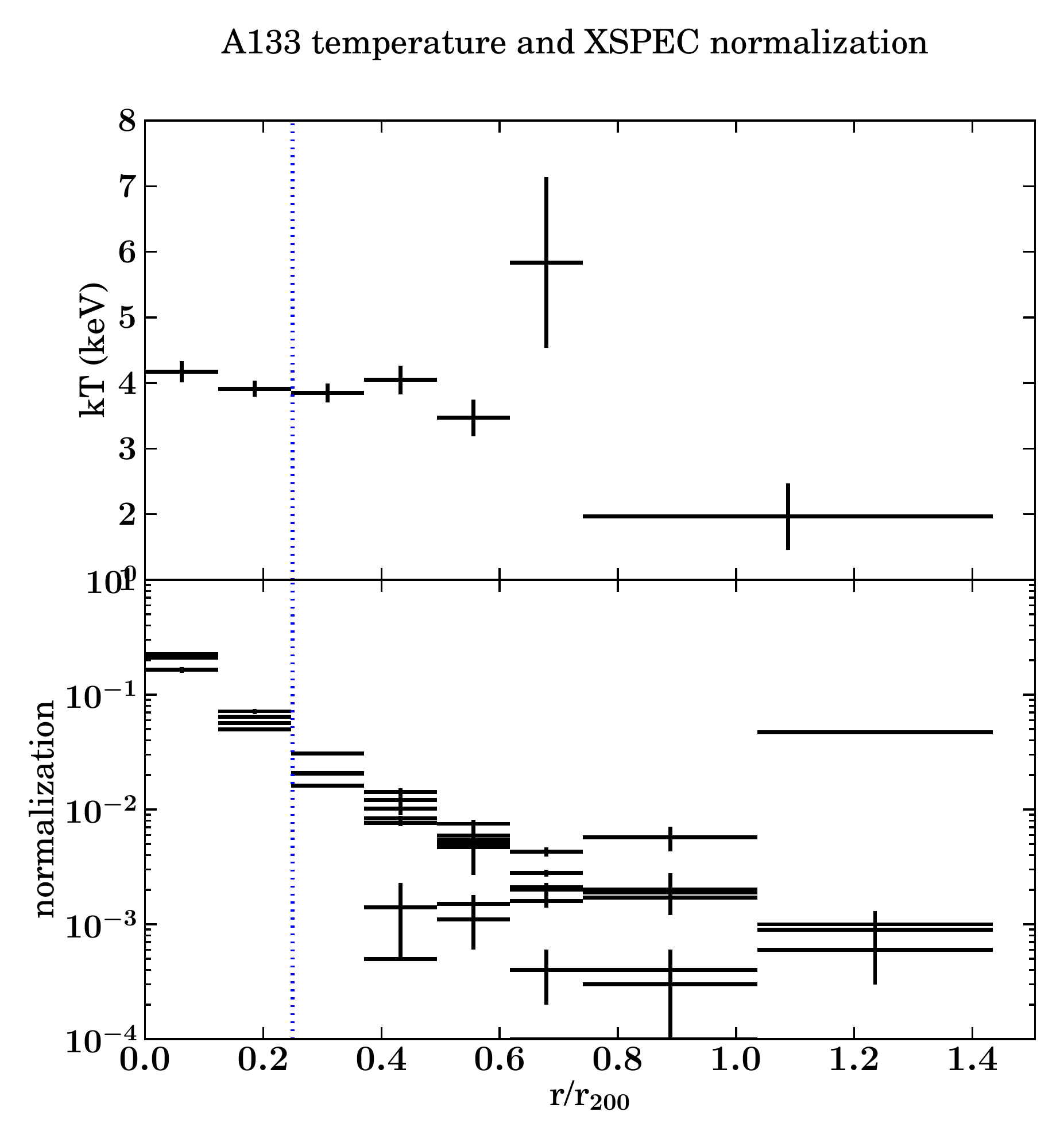}
\end{minipage}
\begin{minipage}{.49\textwidth}
\includegraphics[width=\textwidth]{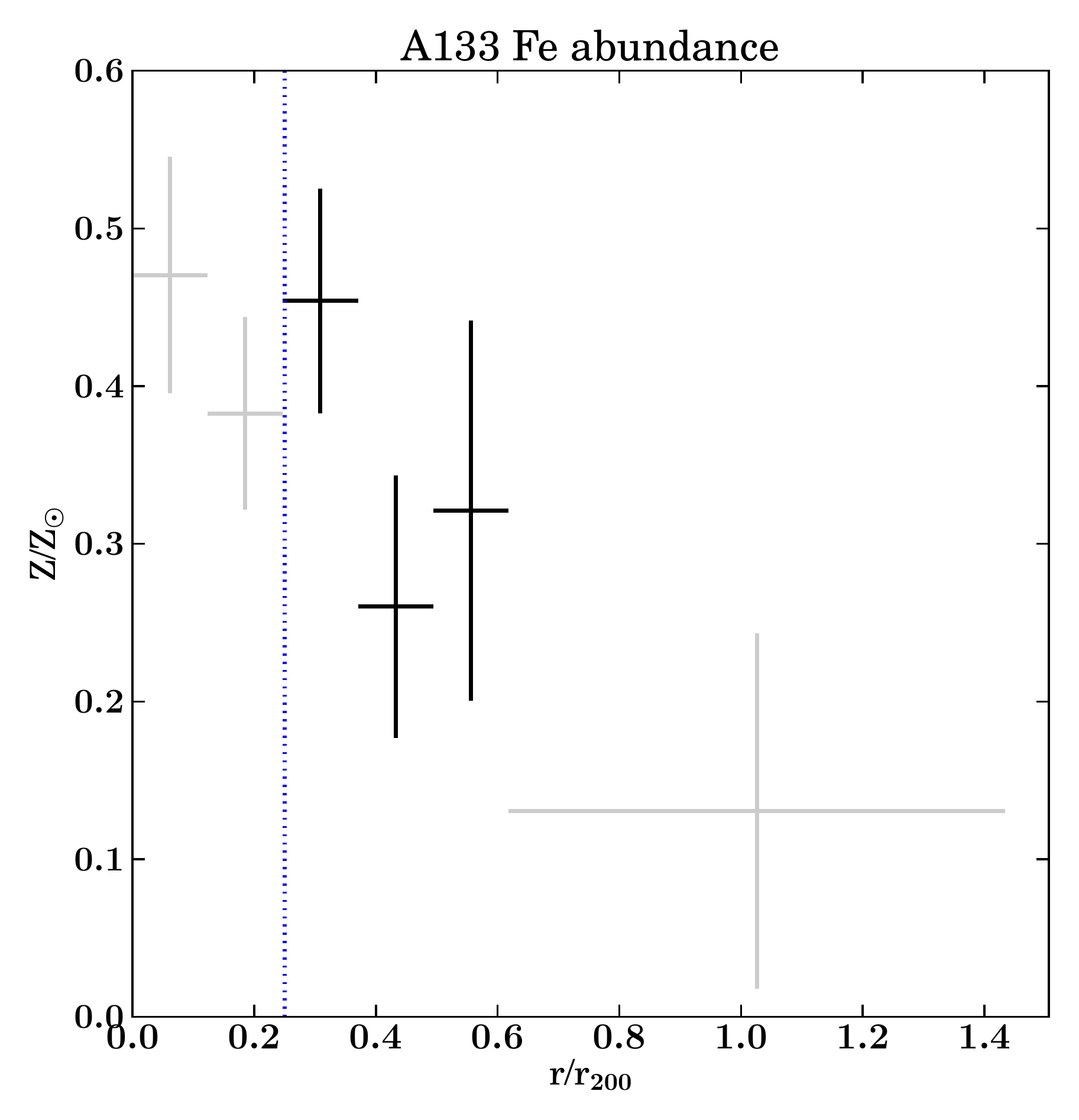}
\end{minipage}
\caption{Same as Fig.~\ref{fig:a262portrait}, but for A~133.}
\label{fig:a133portrait}
\end{figure*}

\begin{figure*}
\begin{minipage}{.49\textwidth}
\includegraphics[width=\textwidth]{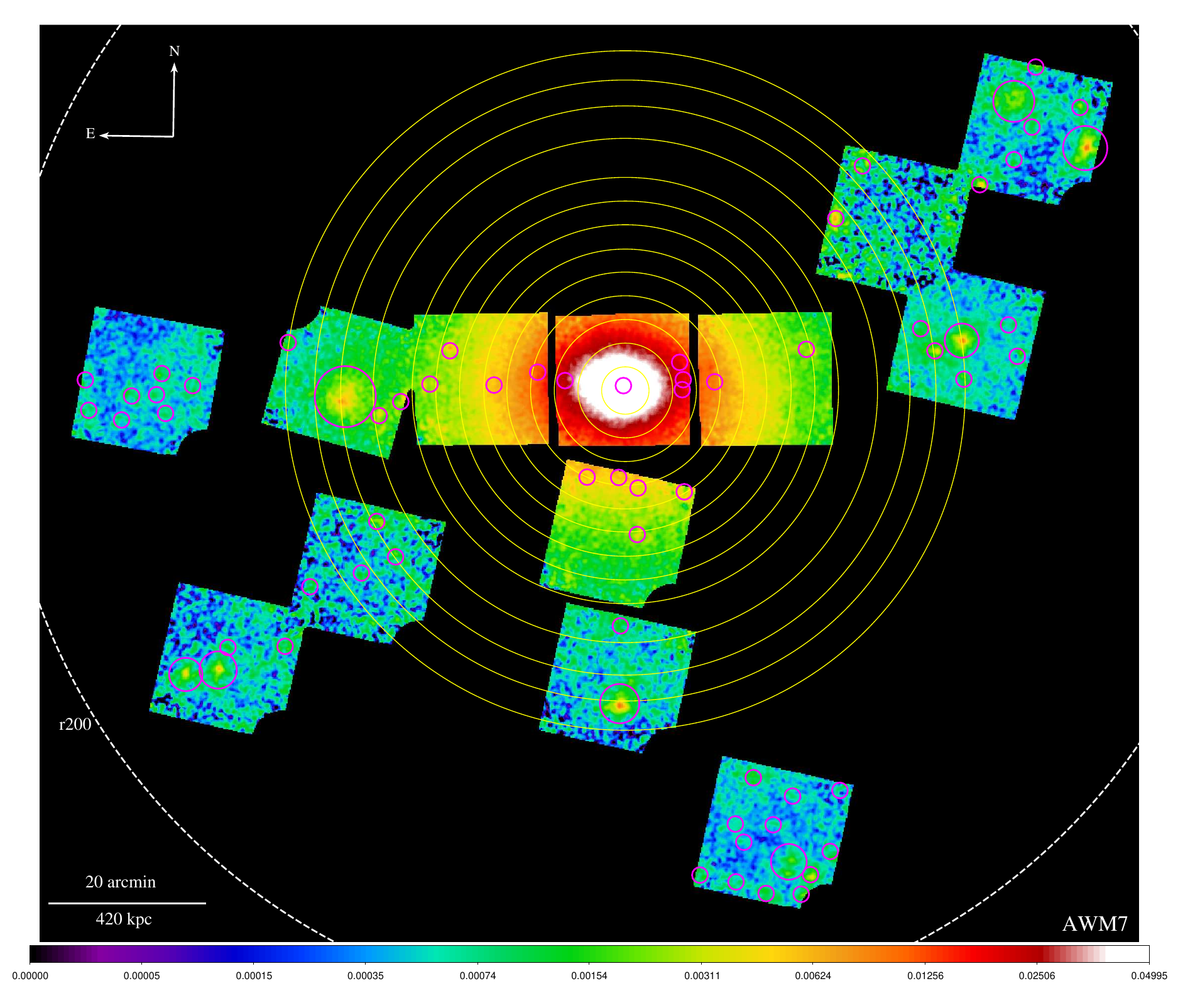}
\end{minipage}
\begin{minipage}{.49\textwidth}
\includegraphics[width=\textwidth]{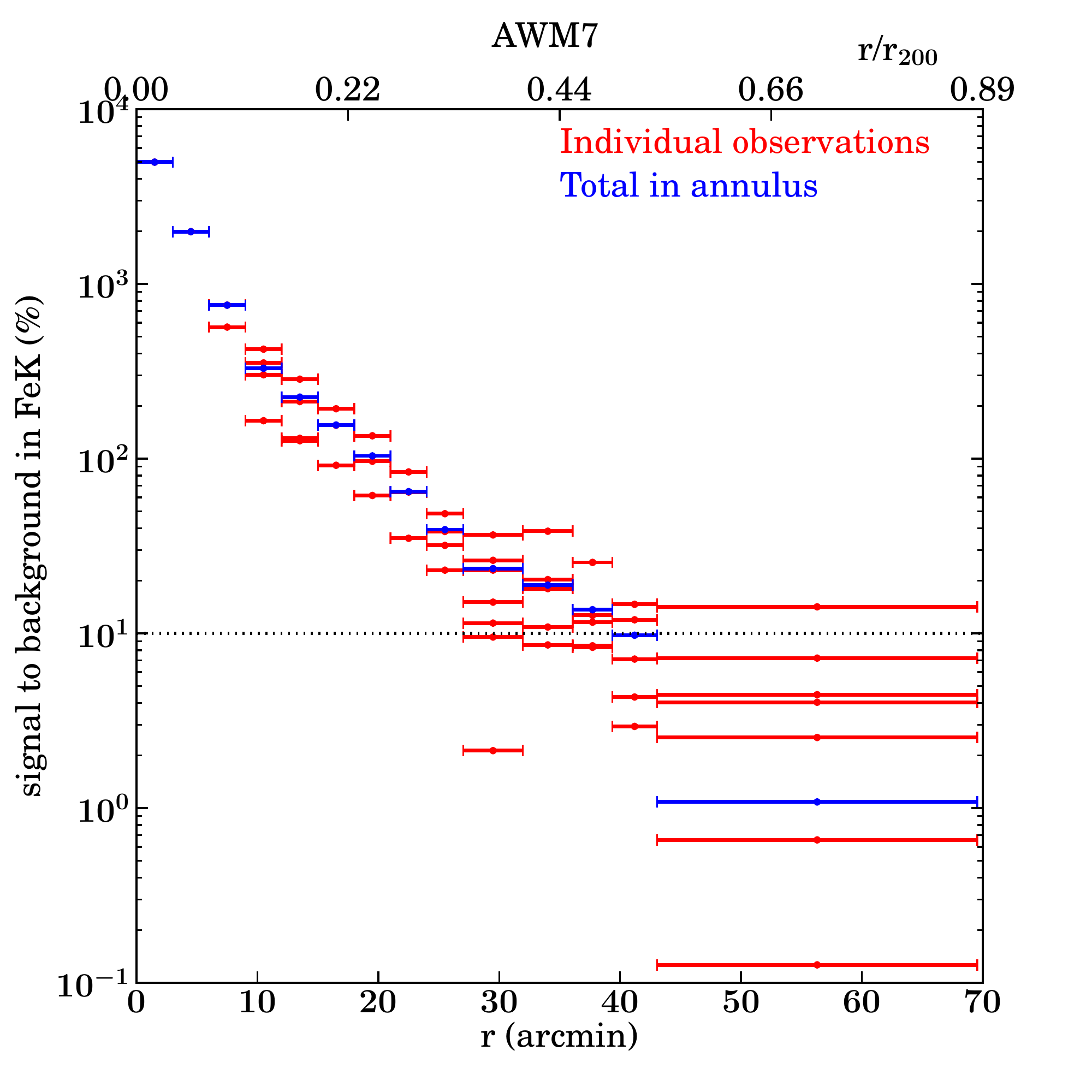}
\end{minipage}
\begin{minipage}{.49\textwidth}
\includegraphics[width=\textwidth]{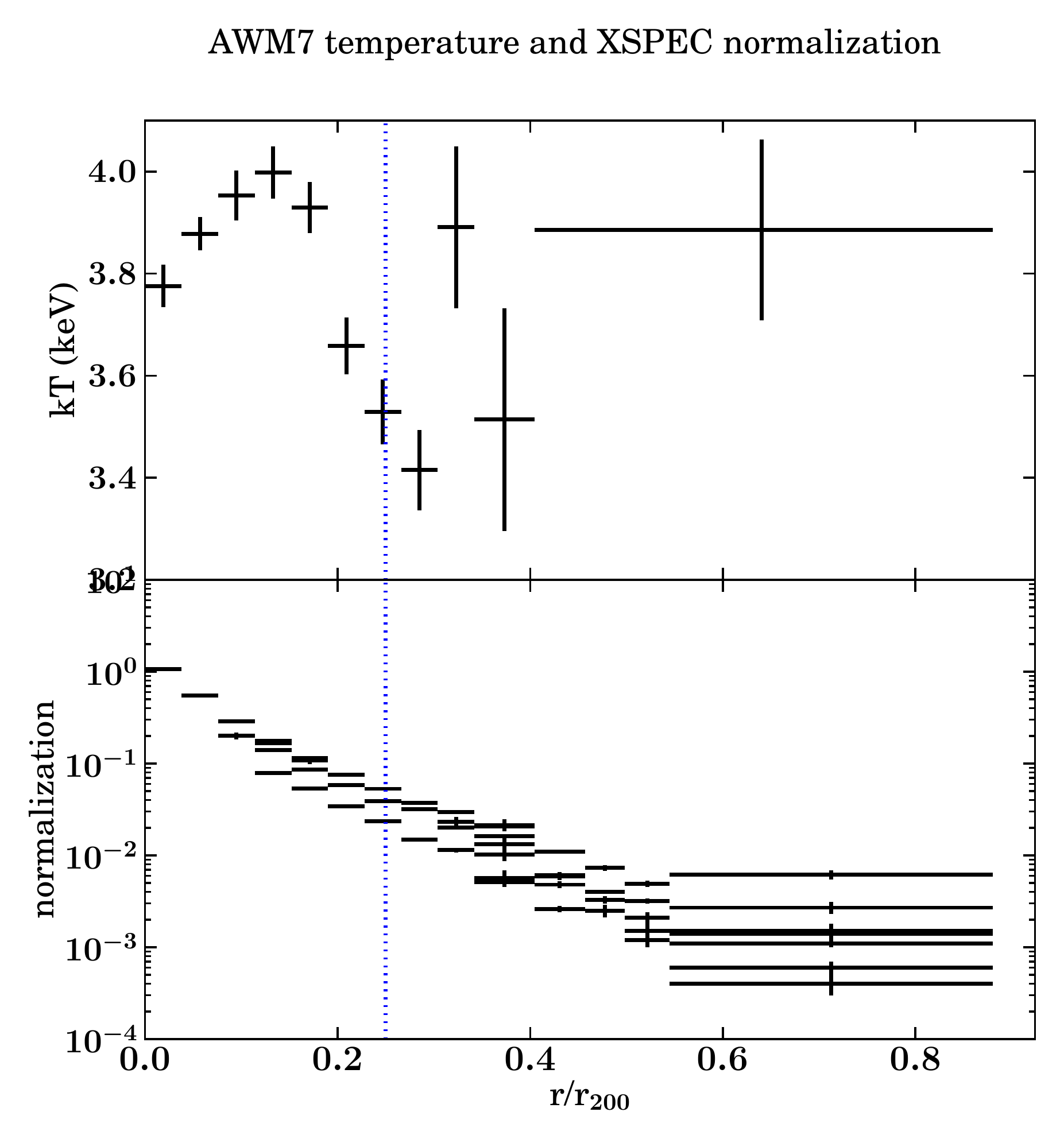}
\end{minipage}
\begin{minipage}{.49\textwidth}
\includegraphics[width=\textwidth]{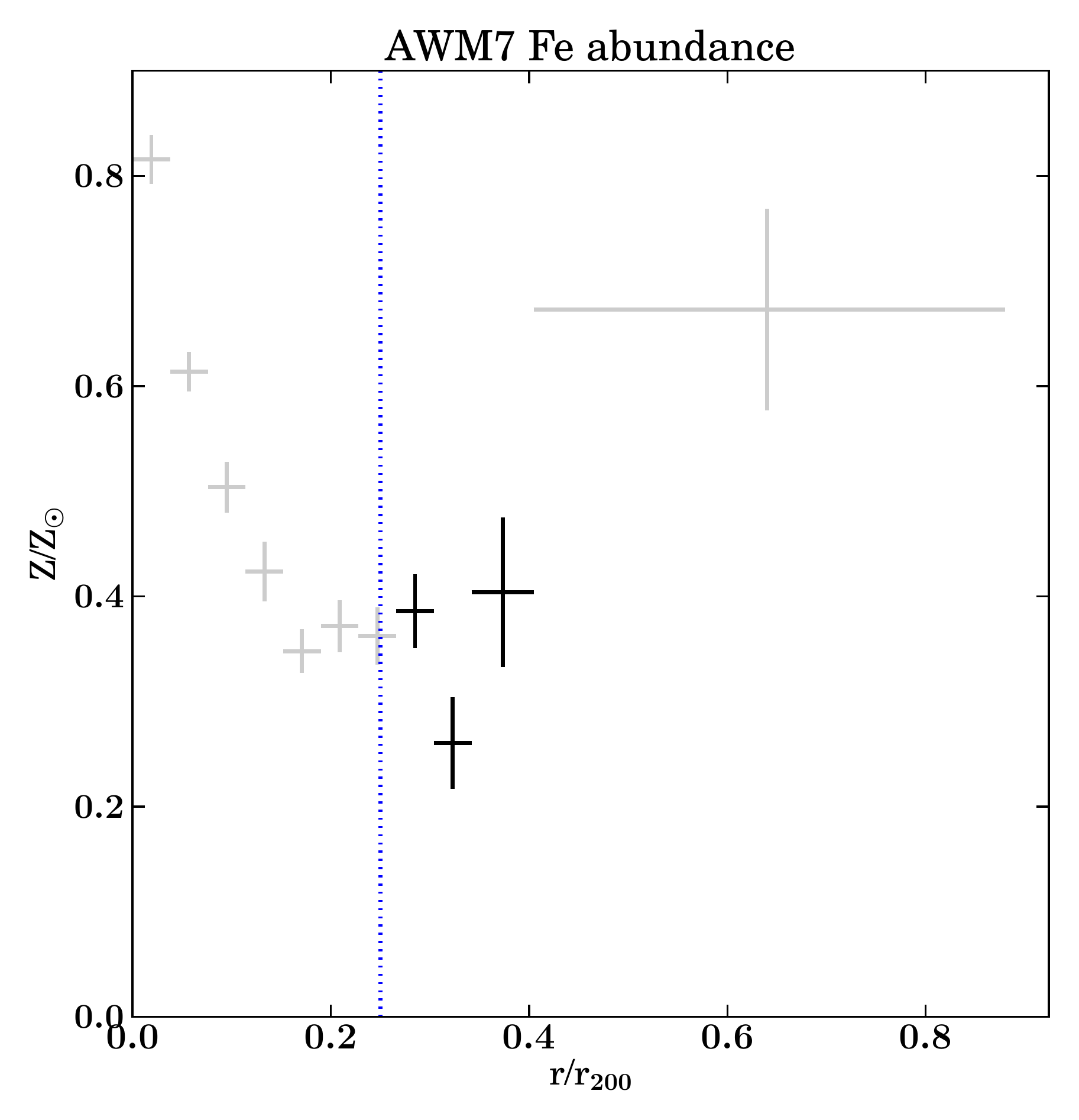}
\end{minipage}
\caption{Same as Fig.~\ref{fig:a262portrait}, but for AWM7.}
\label{fig:awm7portrait}
\end{figure*}

\end{document}